%% file: main.tex
\documentclass[11pt,a4paper]{article}

\usepackage[a4paper, left=2.6cm, right=2.6cm, top=2.7cm, bottom=2.6cm, headsep=0.9cm, headheight=24pt]{geometry}
\usepackage[T1]{fontenc}
\usepackage[utf8]{inputenc}
\usepackage{times}
\usepackage[scaled=0.92]{helvet}
\usepackage{inconsolata}
\usepackage{microtype}
\PassOptionsToPackage{table}{xcolor}
\usepackage{xcolor}
\definecolor{accent}{HTML}{000F46}
\definecolor{accentlink}{HTML}{1D4E9E}
\definecolor{muted}{HTML}{6B7280}
\definecolor{ink}{HTML}{1C2333}
\definecolor{ours}{HTML}{000F46}
\definecolor{oursbg}{HTML}{E3E7F1}
\usepackage{graphicx, booktabs, multirow, tabularx, array, amsmath, amssymb, float, adjustbox, siunitx, latexsym}
\usepackage{fontawesome5}
\usepackage[skins]{tcolorbox}
\usepackage{enumitem}
\setlist{itemsep=2pt, topsep=4pt}
\usepackage{titlesec}
\titleformat{\section}{\sffamily\Large\bfseries}{\textcolor{accent}{\thesection}}{0.6em}{}
\titleformat{\subsection}{\sffamily\large\bfseries}{\textcolor{accent}{\thesubsection}}{0.6em}{}
\titleformat{\paragraph}[runin]{\sffamily\bfseries}{}{0pt}{}[.]
\titlespacing*{\section}{0pt}{16pt}{8pt}
\titlespacing*{\subsection}{0pt}{12pt}{6pt}
\titlespacing*{\paragraph}{0pt}{6pt}{0.6em}
\usepackage[font=small, labelfont={bf,sf,color=accent}, labelsep=period, justification=justified,
            singlelinecheck=false]{caption}
\usepackage{fancyhdr}
\renewcommand{\headrulewidth}{0.4pt}
\renewcommand{\headrule}{\hbox to\headwidth{\color{black!15}\leaders\hrule height \headrulewidth\hfill}}
\fancypagestyle{plain}{\fancyhf{}\fancyfoot[C]{\footnotesize\thepage}\renewcommand{\headrulewidth}{0pt}%
  \fancyhead[L]{\raisebox{-1.5cm}[0pt][0pt]{\includegraphics[height=2.1cm]{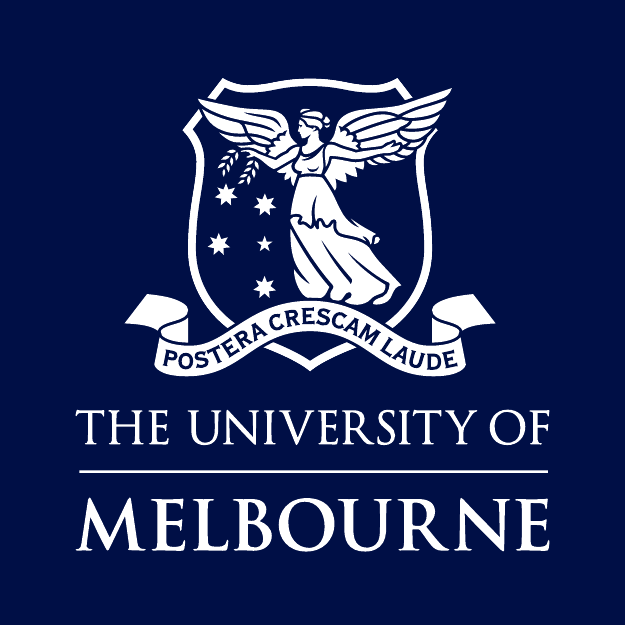}}}}
\usepackage[numbers, sort&compress]{natbib}
\usepackage[hidelinks]{hyperref}
\hypersetup{colorlinks=true, linkcolor=accentlink, citecolor=accentlink, urlcolor=accentlink,
            pdftitle={VeriSpoof: Beyond Real or Fake with One Audio LLM for Speech Deepfake Detection, Source
            Verification, and Grounded Reasoning}, pdfauthor={Yang Xiao}, pdfsubject={Technical report}}
\expandafter\def\expandafter\UrlBreaks\expandafter{\UrlBreaks\do\-}
\usepackage{tocloft}



\newcommand{\name}{VeriSpoof}
\newtcolorbox{figcardbox}[1][\linewidth]{enhanced, width=#1, colback=white, colframe=black!25, boxrule=0.5pt, arc=2pt,
  left=5pt, right=5pt, top=5pt, bottom=5pt, halign=center, drop fuzzy shadow southeast=black!60}
\newcommand{\figcard}[2][\linewidth]{\begin{figcardbox}[#1]\includegraphics[width=\linewidth]{#2}\end{figcardbox}}

\newcommand{\statbox}[2]{%
  \begin{tcolorbox}[enhanced, colback=black!3, colframe=black!12, boxrule=0.4pt, arc=2pt, left=6pt, right=4pt,
                    top=5pt, bottom=5pt, width=\linewidth, height=2.1cm, valign=top]
    \raggedright{\sffamily\bfseries\LARGE\color{accent}#1}\par\vspace{2pt}
    {\sffamily\scriptsize\color{black!75}#2\par}
  \end{tcolorbox}}

\usepackage{soul, xcolor}

\begin{document}

\thispagestyle{plain}
\begin{center}
  {\sffamily\small\bfseries\color{accent}TECHNICAL REPORT \quad\textcolor{muted}{$\bullet$}\quad \textcolor{muted}{November 2026}}\par
  \vspace{14pt}
  {\sffamily\bfseries\fontsize{22}{27}\selectfont VeriSpoof: Beyond Real or Fake\\with One Audio LLM\par}
  \vspace{9pt}
  {\sffamily\normalsize\color{muted} One open audio language model for speech deepfake detection,\\source verification
  and grounded reasoning\par}
  \vspace{14pt}
  {\sffamily\normalsize Yang Xiao, Ting Dang\par}
  \vspace{3pt}
  {\sffamily\small\color{muted} The University of Melbourne\par}
  \vspace{12pt}
  {\sffamily\small
    \href{https://huggingface.co/AustinXiao}{\faCube\ Models}\quad
    \href{https://huggingface.co/datasets/AustinXiao/VeriSpoof-Bench}{\faDatabase\ Benchmark}\quad
    \href{https://github.com/swagshaw/verispoof}{\faGithub\ Code}\quad
    \href{https://huggingface.co/spaces/AustinXiao/VeriSpoof-demo}{\faPlayCircle\ Demo}\quad
    \href{https://swagshaw.github.io/verispoof/}{\faGlobe\ Project page}\par}
\end{center}
\vspace{6pt}

\input{sections/abstract}
\vspace{4pt}

\noindent
\begin{minipage}[t]{0.24\linewidth}\statbox{2.14\%}{pooled EER on 14 Speech-DF-Arena sets: 3rd of 28, 1st with open weights}\end{minipage}\hfill
\begin{minipage}[t]{0.24\linewidth}\statbox{6.2\%}{EER on same-source pairs from generators unseen in training}\end{minipage}\hfill
\begin{minipage}[t]{0.24\linewidth}\statbox{99.6\%}{PartialEdit fakes caught after reasoning; the one-step answer catches 78.4\%}\end{minipage}\hfill
\begin{minipage}[t]{0.24\linewidth}\statbox{99.5\%}{decisions that follow an edit of the stated evidence never seen in training}\end{minipage}\hfill

\begin{figure}[p]
  \centering
  \figcard[0.74\linewidth]{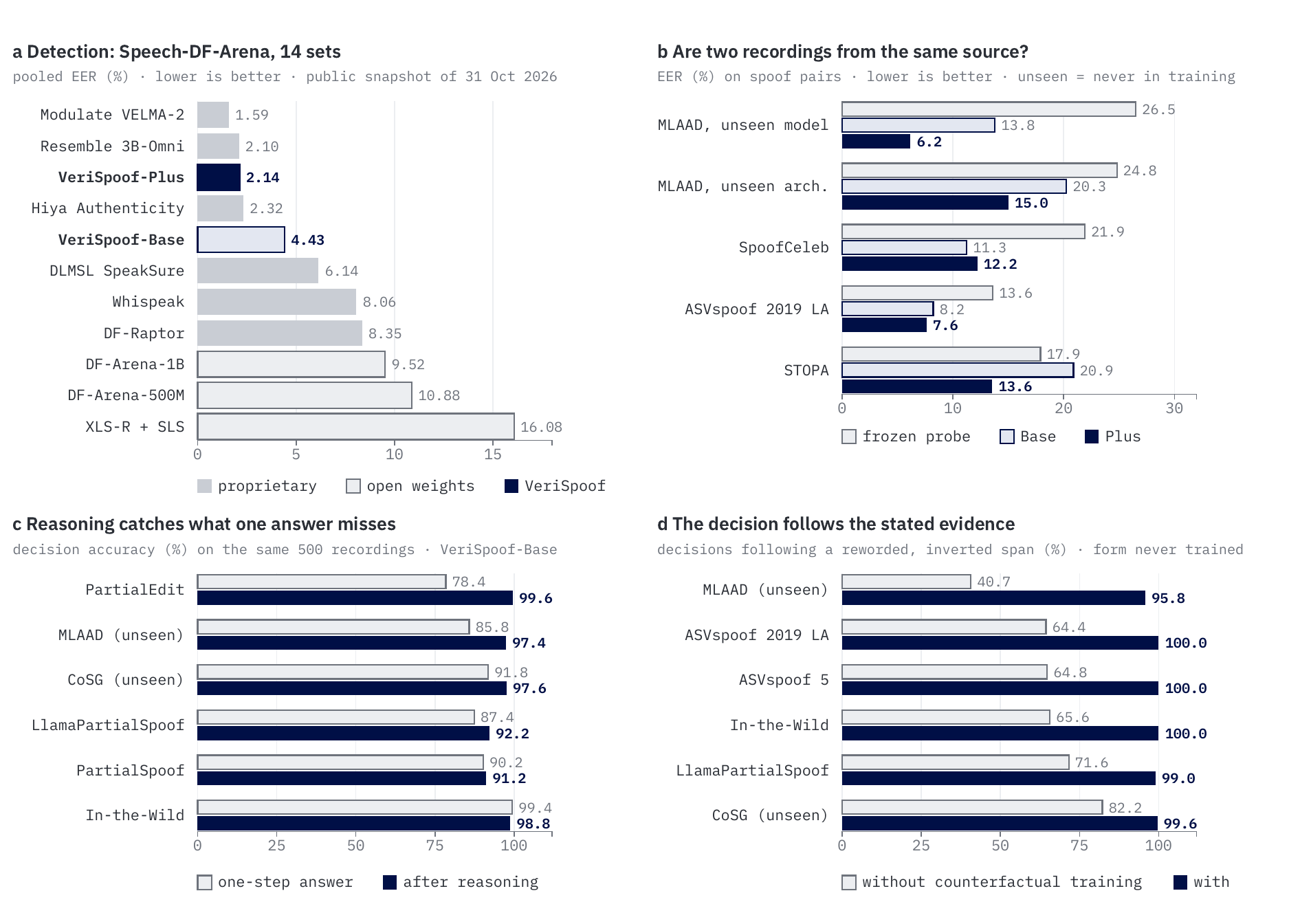}
  \caption{\textbf{\name{} at a glance.} (a) Detection on the 14 test sets of Speech-DF-Arena \citep{dfarena}, pooled
    EER (one threshold over all sets, the board's ranking); the public snapshot of 31 Oct 2026 with both \name{}
    models inserted, top 11 shown. (b) The same weights decide whether two spoofed recordings come from the same
    generator, against a pair probe on the frozen anti-spoofing encoder DF-Arena-1B (\S\ref{sec:res-source}).
    (c) Accuracy of the one-step answer and of the decision \name{}-Base reaches after writing its rationale
    (\S\ref{sec:res-reason}). (d) Share of decisions that follow a localization replaced by its opposite, in a wording
    never used in training (\S\ref{sec:res-faith}).}
  \label{fig:glance}
  \vspace{6pt}
  \begin{minipage}{0.92\linewidth}
    \small
    \tableofcontents
  \end{minipage}
\end{figure}

\input{sections/introduction}

\input{sections/related}
\input{sections/method}
\input{sections/data}
\input{sections/evaluation}
\input{sections/results}
\input{sections/conclusion}

\section*{Acknowledgements}
\addcontentsline{toc}{section}{Acknowledgements}
Experiments used computational resources of the National Computational Infrastructure (NCI Australia) and of the
University of Melbourne. Generative AI tools were used to assist with writing code, running and summarising
experiments, and drafting and editing this report; the authors reviewed and take responsibility for all content.

\bibliographystyle{unsrtnat}
\bibliography{refs}

\clearpage
\appendix
\input{sections/appendix}

\end{document}

%% file: sections/abstract.tex
\begin{tcolorbox}[enhanced, colback=accent!5, colframe=accent, boxrule=0pt, leftrule=2.5pt, arc=0pt, sharp corners,
                  left=8pt, right=8pt, top=6pt, bottom=6pt]
{\sffamily\bfseries\small\color{accent}Abstract}\par\vspace{3pt}
\small
A speech deepfake detector reports how likely a recording is synthetic, but a forensic analyst also asks which system
produced it, whether two recordings share a source, and on what evidence the conclusion rests. We present \textit{\name{}}, an audio language model that unifies five forensic tasks — deepfake detection, generator attribution (generator, acoustic model and vocoder), recording-condition analysis, synthesized-span localisation and source verification — with one set of weights, and supports each decision with grounded rationales whose every statement can be independently verified. Every answer is a sentence whose
likelihood is read out as a score, so detection and source verification need no separate heads and can be thresholded
at an operating point. Every explanation consists of statements with an independent check: observations are measured
on the waveform the model hears, synthesized spans come from labels, and the source profile from generator metadata.
Counterfactual examples teach the decision to follow the stated evidence. On the 14 test sets of Speech-DF-Arena,
\textit{\name{}-Plus} reaches 2.14\% pooled EER, third of 28 systems and first among open-weight ones, and \textit{\name{}-Base},
trained on the official training splits of eight corpora only, reaches 4.43\%, less than half the EER of the next
open-weight detector. With the same weights, \textit{\name{}-Plus} decides whether two recordings from generators unseen in
training share a source at 6.2\% EER. Controlled studies show that the audio representation decides what is
detected, the LLM how well two recordings are compared, and that source supervision is necessary for neural-codec
fakes. Rationales can be checked statement by statement, catch partial fakes that the one-step answer misses (99.6
against 78.4\% on PartialEdit), and determine the decision under edits never seen in training (99.5\%). Only 29.7\,M of the 2.85\,B parameters are
trained, so training fits a single 24\,GB GPU with gradient checkpointing. Both models, the inference code, a demo, and
\textit{VeriSpoof-Bench}, the evaluation suite of all these questions, are released.
\end{tcolorbox}

%% file: sections/introduction.tex
\section{Introduction}
\label{sec:intro}

Speech deepfakes pose a significant security threat nowadays. A suspicious speech recording can raise several related questions: whether it is synthetic, which system produced it, and whether another recording came from the same source. Consider a voice message submitted as evidence. A deepfake detector can tell an analyst that it is probably synthetic,
but the analyst's next questions are harder: which system produced it, whether it came from the same system as another
message in the case, and on what grounds the detector believes so. The answers link a recording to a tool, and through
the tool to a person or a campaign; a conclusion that cannot be explained cannot be contested. 

Research has made the first question tractable: detectors trained on large and diverse corpora reach low error rates
across many benchmarks \citep{doss, teffic, dfarena}. Source attribution, verification, and explanation have also been studied. For a forensic workflow, open detectors leave three problems.
\begin{enumerate}[label=\textbf{\arabic*.}, leftmargin=*]
  \item \textbf{One score answers one question.} Source tracing assigns a recording to a closed set of known systems or
        attributes \citep{falez2025sourcetracing, stopa}, while comparing two recordings needs a separate Siamese network
        \citep{forensicsim, audity}. Each further question costs another model, trained on its own data.
  \item \textbf{Explanations cannot serve as evidence.} Audio language models (ALMs) for deepfakes add explanations
        \citep{holiantispoof, thinkomni, iclad}, but these are free text written or distilled by a teacher model.
        Whether a stated observation is true of the recording is not checked, and whether the decision depends on it
        is not tested. Such an explanation can be fluent and wrong, or correct and irrelevant to the decision
        \citep{turpin2023unfaithful, lanham2023faithfulness}.
  \item \textbf{It is unclear where to invest.} Speech-recognition encoders have been argued to discard the artifacts
        of synthesis \citep{alethia}, and source labels are often assumed to help detection \citep{holiantispoof}
        without controlling for the number of authenticity labels or training steps. Which part of an ALM limits detection,
        and which limits comparing recordings, has not been measured.
\end{enumerate}

\begin{figure}[t]
\centering
\figcard{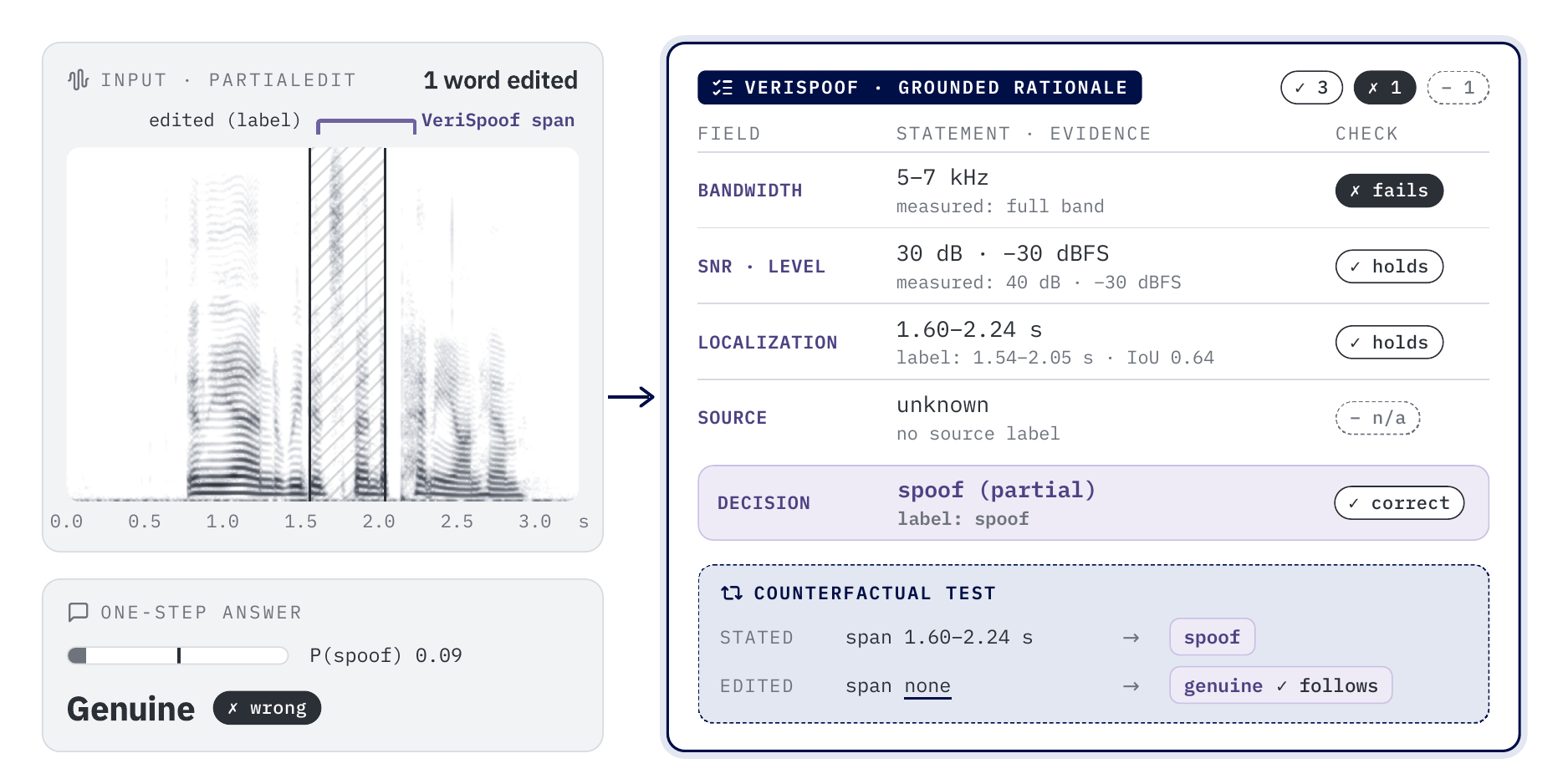}
\caption{\textbf{One score versus checkable evidence} on the same recording, in which one word of genuine speech was
replaced by synthesized speech. Asked directly, the model judges the recording genuine (left). Asked to reason,
\textit{\name{}-Base} states observations, a synthesized span and a source profile before deciding (right). Each statement is
checked against a waveform measurement or a label ($\checkmark$ holds, $\times$ fails, -- no label), and replacing the
stated span by ``none'' flips the decision. Statements are abridged; the full output is in
Figure~\ref{fig:app-examples}.}
\vspace{-5mm}
\label{fig:teaser}
\end{figure}

VeriSpoof addresses all three problems in a single model: it unifies detection, attribution, recording-condition analysis, localisation and source verification, and accompanies every decision with rationales that can be checked statement by statement against the signal. For the first, it casts every forensic question as a question with candidate answers
written as sentences and reads the likelihood of each answer as a score, so one model detects spoofs, names the
generator, its acoustic model and vocoder, describes the recording channel and verifies whether two recordings share a
source (Figure~\ref{fig:overview}). For the second, it writes rationales in fixed fields, each with a target computed from 
the exact audio the model hears, so that every statement has a check without trusting the model. 
Training on counterfactual rationales further ensures that the decision follows the stated evidence
(Figure~\ref{fig:teaser}). 
For the third, we run controlled studies that change one component or one form of supervision at a time on the same test trials, measuring which part of the model limits detection and which limits source comparison.

\begin{tcolorbox}[enhanced, colback=white, colframe=accent, boxrule=0.6pt, arc=2pt, left=8pt, right=8pt, top=6pt,
                  bottom=6pt, title={\sffamily\bfseries\small Key results}, colbacktitle=accent, coltitle=white,
                  fonttitle=\sffamily\bfseries\small]
\small
\begin{itemize}[label={\color{accent}$\blacktriangleright$}, leftmargin=1.4em, itemsep=4pt]
  \item \textbf{Best open-weight detector.} On the 14 Speech-DF-Arena sets, \textit{\name{}-Plus} reaches 2.14\% pooled and
        1.79\% mean EER (third of 28 systems on the public snapshot), and \textit{\name{}-Base}, trained on 0.38\,M utterances
        of official training splits, 4.43\%; the best other open-weight system reaches 9.52\%
        (\S\ref{sec:res-detection}).
  \item \textbf{The same weights verify sources.} For pairs from generators unseen in training, \textit{\name{}-Plus} decides
        whether they share a source at 6.2\% EER for new models of known architectures and 15.0\% for new
        architectures, against 26.5 and 24.8\% for a pair probe on a frozen anti-spoofing encoder; general audio LLMs
        answer near chance (\S\ref{sec:res-source}).
  \item \textbf{Representation decides detection, the LLM decides comparison.} With the encoder fixed, how its
        features are used lowers ASVspoof~5 EER from 9.6 to 3.1\% while source verification stays at 16--19\%;
        swapping the LLM moves detection by 0.5 points and source verification by 5.5 (\S\ref{sec:res-perception}).
  \item \textbf{Source labels are necessary for neural-codec fakes.} At a matched number of authenticity labels,
        source supervision helps on 8 of 16 sets and hurts on one; with five times more authenticity labels instead,
        CodecFake+ EER stays at 12.6 against 2.2\% (\S\ref{sec:res-supervision}).
  \item \textbf{Checkable, useful and faithful rationales.} Statements are checked one by one (SNR and level correct
        for 88--100\% of recordings); the decision after reasoning catches partial and unfamiliar fakes the one-step
        answer misses (99.6 vs.\ 78.4\% on PartialEdit); and with counterfactual training it follows a reworded,
        inverted localization in 99.5\% of cases, against 73.8\% without (\S\ref{sec:res-rationale}--\ref{sec:res-faith}).
  \item \textbf{Light to train and run.} Only 29.7\,M of the 2.85\,B parameters are trained. With gradient checkpointing, every
        training stage fits a single 24\,GB GPU, and we estimate that training fits a 12\,GB GPU and inference an 8\,GB
        GPU (\S\ref{sec:res-speed}).
\end{itemize}
\end{tcolorbox}

Both models are released: \textit{\name{}-Base}, trained on the official training splits of eight corpora, and \textit{\name{}-Plus}, which extends Base with twelve additional corpora and a larger balanced training mixture (\S\ref{sec:data}), together with the inference code with a vLLM \citep{vllm} backend and a hosted demo. We also release \textit{\name{}-Bench}, the evaluation suite of this report: 54 test subsets with 177,852 items covering detection, attribution, source verification, recording conditions, grounded reasoning, and faithfulness, with scoring scripts and the per-item outputs of both models.

%% file: sections/related.tex
\section{Related work}
\label{sec:related}

\paragraph{Speech deepfake detection} Most detectors pair a self-supervised front-end with a light back-end
\citep{aasist, xlsrsls, nes2net}. Recent work improves generalization through spectral residuals \citep{sonar},
high-order feature interactions \citep{hyperpotter}, style--linguistics mismatch \citep{slim}, dedicated pre-training
\citep{alethia}, or data composition at scale \citep{doss, teffic}. Evaluation has moved from ASVspoof
\citep{asvspoof2019, asvspoof5} and In-the-Wild \citep{itw} to multi-set boards such as Speech-DF-Arena
\citep{dfarena} and SpoofRadar \citep{spoofradar}. These systems return one score per recording; \textit{\name{}} returns its
detection score as the likelihood of its own answer and is evaluated on both boards.

\paragraph{Source tracing and verification} Source tracing assigns a recording to a known system or to attributes such
as its acoustic model and vocoder \citep{falez2025sourcetracing, stopa}, on corpora with many generators such as
MLAAD \citep{mlaad} and SpoofCeleb \citep{spoofceleb}. Dedicated Siamese or two-branch networks score whether two
recordings come from the same generator \citep{forensicsim, audity}. We pose this comparison to the model that also
detects and explains, on open-set protocols with unseen generators or architectures.

\paragraph{Audio language models for deepfakes} HoliAntiSpoof unifies detection, method recognition and localization
\citep{holiantispoof}, ThinkOmni generates reasoning with localization \citep{thinkomni}, and ICLAD prompts an ALM with
comparison examples \citep{iclad}. However, general-purpose ALMs \citep{qwen25omni, qwen3omni, audioflamingo3, kimiaudio} are not trained for these tasks. Their
rationales, where present, are free text from a teacher model and are not checked against the signal: ICLAD reports
hallucinated explanations, and ThinkOmni's training rationales are written by a model shown the labels. SpeakerLLM
\citep{nam2026speakerllm} unifies speaker profiling, recording conditions and verification reasoning for speakers.

\paragraph{Faithful explanations} In NLP, an explanation is faithful when the decision depends on it, which is tested
by editing the rationale and observing the decision \citep{atanasova2023faithfulness, lanham2023faithfulness,
turpin2023unfaithful}. We apply such tests to audio, with edit forms held out from training, and train for them with
counterfactual rationales.

%% file: sections/method.tex
\section{\name{}}
\label{sec:method}

\begin{figure}[t]
\centering
\figcard{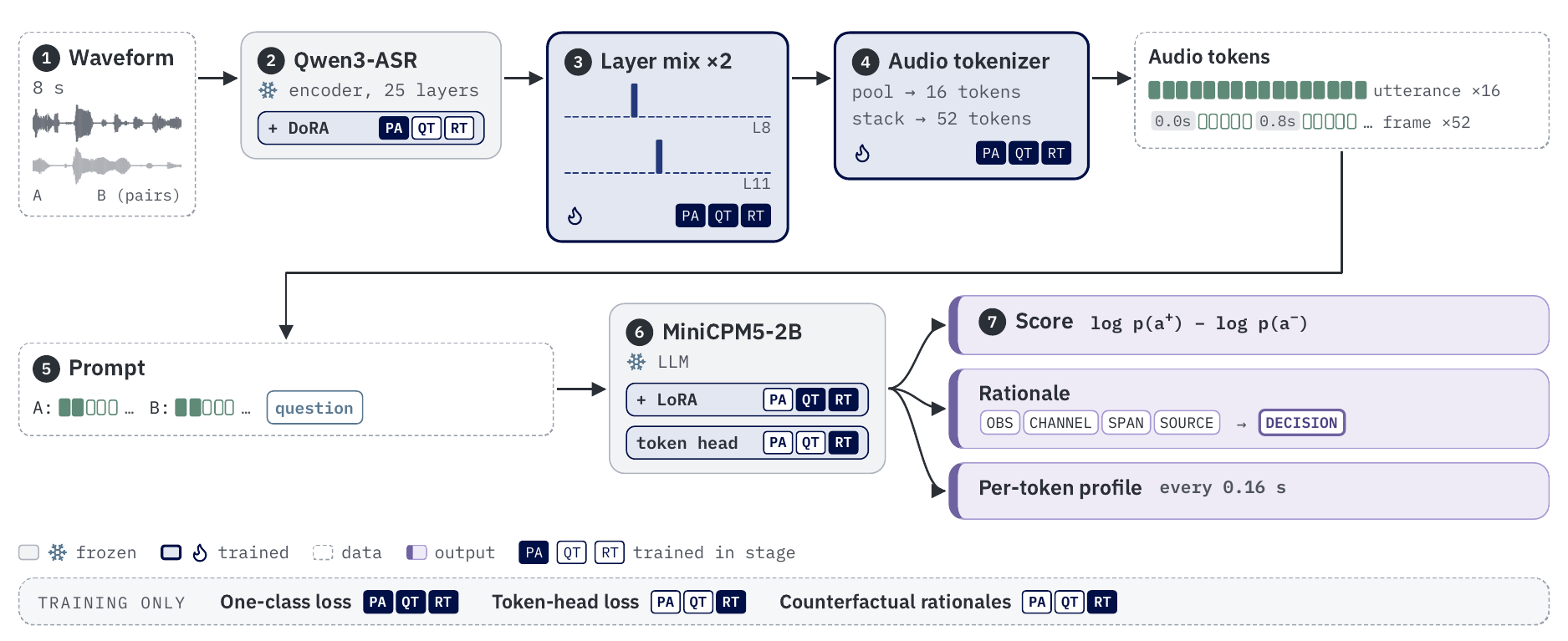}
\caption{\textbf{\name{}.} (1) An 8\,s waveform, or two for a pair, enters (2) the Qwen3-ASR
encoder, frozen apart from DoRA adapters. (3) Two learned mixes over its 25 layers feed (4) the audio tokenizer, which pools the recording into 16 utterance
tokens and stacks pairs of frames with a text time marker every 0.8\,s. (5) The prompt
places the tokens of A, then those of B, before the question; (6) the frozen LLM with LoRA reads it and (7) returns a
score, the log-likelihood difference of two answer sentences, a rationale whose decision comes last, or a per-token
synthetic-speech profile. Grey: frozen; dark blue: trained by us, with the stages that train each part (PA: perception
alignment, QT: forensic question tuning, RT: grounded rationale tuning).}
\label{fig:overview}
\end{figure}

\subsection{Problem Formulation: Forensic Questions as Answer Scoring}
\label{sec:formulation}
Let $u$ denote the input, either a single recording $x$ or a pair $(x, x')$, 
and $q$ a question with a set of candidate answers $\mathcal{A}_q$ each written as
sentences. An ALM with parameters $\theta$ assigns each answer $a \in \mathcal{A}_q$, the likelihood $p_\theta(a \mid u, q)$ of its tokens. 
 For a binary yes--no question, let 
$a^{+}$ and $a^{-}$ denote the positive and negative answers; we read the score, 
\begin{equation}
s_q(u) = {} \log p_\theta(a^+ \mid u,q)
             - \log p_\theta(a^- \mid u,q).
\label{eq:score}
\end{equation}
which can be evaluated by equal error rate (EER) and thresholded at a chosen false-alarm rate. Questions with more
answers are scored by the accuracy of the most likely one. \textit{\name{}} answers four families of questions
(Figure~\ref{fig:tasks}): \emph{authenticity}; \emph{attribution} of the generator, acoustic model or vocoder;
\emph{recording condition} (noise, reverberation, codec, post-processing); and \emph{source verification} on pairs
(same generator, acoustic model or vocoder?). No question has its own output head. For detection, for example, the
question is ``Is this recording genuine human speech or spoofed speech? Options: bona fide; spoof.'' and the two
answers are full sentences (``This recording is a spoof: it is synthetic or converted speech, not genuine human
speech.''); the released models label a recording spoofed when $s_q(x) > 0$. Figure~\ref{fig:questions} shows
the template of every question with an example answer.

Source verification is not the same as comparing attributes: two recordings from different systems can share every attribute class (e.g., generator family, acoustic model and vocoder), so a model that answers ``same source'' whenever
the attributes agree fails on such pairs. Source-verification test sets therefore balance three strata: SAME,
REVERSAL (different systems, every attribute class shared) and DIFF, so that attribute agreement alone does not
reveal the answer.

\subsection{Model}
\label{sec:model}
\paragraph{Perception in two resolutions} A Qwen3-ASR encoder \citep{qwen3asr} reads an 8\,s window of the
recording. A learned weighted sum of the hidden states 
of all 25 of its layers is passed to an audio tokenizer with two outputs: 16 utterance tokens,
which carry what is constant across the recording and serve attribution and verification, and 52 frame tokens of
0.16\,s, which preserve temporal information. Time markers every 0.8\,s are inserted between frame tokens as text, so that the LLM can
state spans in seconds. The encoder is adapted with DoRA \citep{dora} ($r{=}16$) during the first training stage only and frozen afterwards. 
For a pair, the tokens of both recordings are concatenated in the prompt.

\paragraph{LLM and read-out} The LLM is MiniCPM5-2B \citep{minicpm5, minicpm4}, adapted with LoRA \citep{lora} ($r{=}16$,
$\alpha{=}32$) on the attention projections. The model has 2.85\,B parameters, of which 29.7\,M are trained (\S\ref{sec:res-speed}). The model is trained with 
next-token prediction on the answer tokens and answers are scored by Eq.~\eqref{eq:score}. Two auxiliary losses support perception: a one-class
authenticity loss \citep[OC-softmax;][]{ocsoftmax} on the pooled audio tokens, which pulls bona fide speech towards a
learned centre, and a per-token head that classifies each frame token as genuine or synthesized from the LLM's hidden
state. The per-token head gives a synthetic-speech profile over time at no extra cost and serves as a reference for
how much localization information the hidden states carry (\S\ref{sec:res-rationale}). The model has two output modes: a one-step answer scored by Eq. (\ref{eq:score}), and a rationale followed by a decision, described in \S\ref{sec:rationales}.

\subsection{Rationales written from what the model hears}
\label{sec:rationales}
For forensic use, a decision must come with evidence that can be verified and contested, so \textit{\name{}} explains its decisions with rationales whose every statement can be checked independently of the model. A rationale $r = (r_1, \dots, r_k; d)$ is a sequence of statements in fixed fields followed by a decision $d$. Each
statement type $\tau$ has a checker $c_\tau(x, r_i) \in \{\text{true}, \text{false}\}$ that does not involve the model:
\begin{itemize}[leftmargin=*]
  \item \textsc{observations} (bandwidth class, SNR, speech level, clipping, digital silence) are checked against
        measurements on the waveform;
  \item \textsc{channel\_status} (noise, reverberation, codec) against the logged recording condition;
  \item \textsc{localization} (synthesized spans in seconds, or none) against labelled spans;
  \item \textsc{source\_profile} (generation method, acoustic modelling, waveform generation) against generator
        metadata.
\end{itemize}
A rationale is \emph{grounded} if every statement has a checker, so its correctness can be measured statement by
statement. Channel statements describe the recording, not its authenticity, and are declared non-evidential. The data loader computes them directly from the exact window the model receives after augmentation: it measures observations on the waveform, takes channel statements from the augmentation log, maps labelled spans to window time, and reads the source profile from generator metadata. Each target is therefore true of its input, and the decision, written last, follows from the localization: a recording is labeled spoofed if and only if at least one synthesized span is stated, and genuine otherwise. The decision $d$ is the final generated token sequence after the rationale fields; it is read as a score by Eq. (\ref{eq:score}) when thresholding is needed, or taken as the generated text directly. Verification rationales have the fields \textsc{channel\_status},
\textsc{fingerprint}, \textsc{profile\_compatibility} and \textsc{decision}; they state the fingerprint verdict before
comparing attributes, because the attribute comparison can mislead the decision on REVERSAL pairs
(\S\ref{sec:res-reason}). Figure~\ref{fig:app-examples} shows one output of each kind in the Appendix.

\subsection{Three training stages}
\label{sec:stages}
Training has three stages, each with its own trainable parts (Table~\ref{tab:app-hparams}).
\begin{enumerate}[label=\textbf{\arabic*.}, leftmargin=*]
  \item \textbf{Perception alignment (PA)} trains the tokenizer and the encoder's DoRA adaptors on detection and source
        questions with the LLM frozen, so that the audio representation itself learns the relevant information before the LLM can compensate for weak audio
        tokens. The representation study of \S\ref{sec:res-perception} is run at this stage for the same reason.
  \item \textbf{Forensic question tuning (QT)} adds LoRA on the LLM and trains it on all perception questions,  a form of multi-task instruction tuning: acoustic model,
        vocoder, source profile, recording condition, and pair questions on generator, acoustic model, vocoder and the
        authenticity of both recordings.
  \item \textbf{Grounded rationale tuning (RT)} introduces detection and verification rationales and counterfactual examples at a lower learning rate. To preserve the one-step answering ability acquired in QT, the question-answering tasks of QT are replayed throughout this stage.
\end{enumerate}
Each question has several paraphrases. Two models share this recipe and differ in their training data
(\S\ref{sec:data}): \textit{\name{}-Base}, the model of all controlled studies, and \textit{\name{}-Plus}. Counterfactual examples and their training procedure are described in \S\ref{sec:counterfactual}.

\subsection{Counterfactual training}
\label{sec:counterfactual}
Counterfactual training is applied in stage 3 (grounded rationale tuning). Grounded rationales alone do not make the decision depend on the rationale: the model can decide from the audio and write
a matching rationale beside it. To make the decision follow the rationale, in 20\% of the rationale examples we give the model a rationale whose localization is
reversed (a labelled span replaced by ``none'', or ``none'' replaced by a span) and train only the decision that
follows. The model is never taught to state something false about the audio. It is taught that the decision must
follow the localization it is given. We further test faithfulness with interventions: a statement $r_i$ is replaced by
$\tilde r_i$ and the model decides again. The decision is faithful if it follows an edited localization (``no
synthesized segment'' $\Rightarrow$ genuine) and remains the same when an edited non-evidential statement is edited. Besides the edit form used
in training, we test held-out forms: a paraphrased localization, an inserted span, and contradicted channel or
observation statements.

\subsection{Inference}
\label{sec:inference}
One-step answers are scored by the likelihood of each candidate answer sentence; rationales are generated greedily.
The released package exposes detection (\texttt{detect}), rationales with the per-token profile (\texttt{explain}),
source verification at the generator, acoustic-model or vocoder level (\texttt{same\_source}), verification
rationales (\texttt{compare}), and closed questions on recording conditions and on generators from a list the user
supplies. A vLLM \citep{vllm} backend generates rationales from the same prompt embeddings (\S\ref{sec:res-speed}).


%% file: sections/data.tex
\section{Training data and recipe}
\label{sec:data}

\begin{figure}[t]
\centering
\figcard{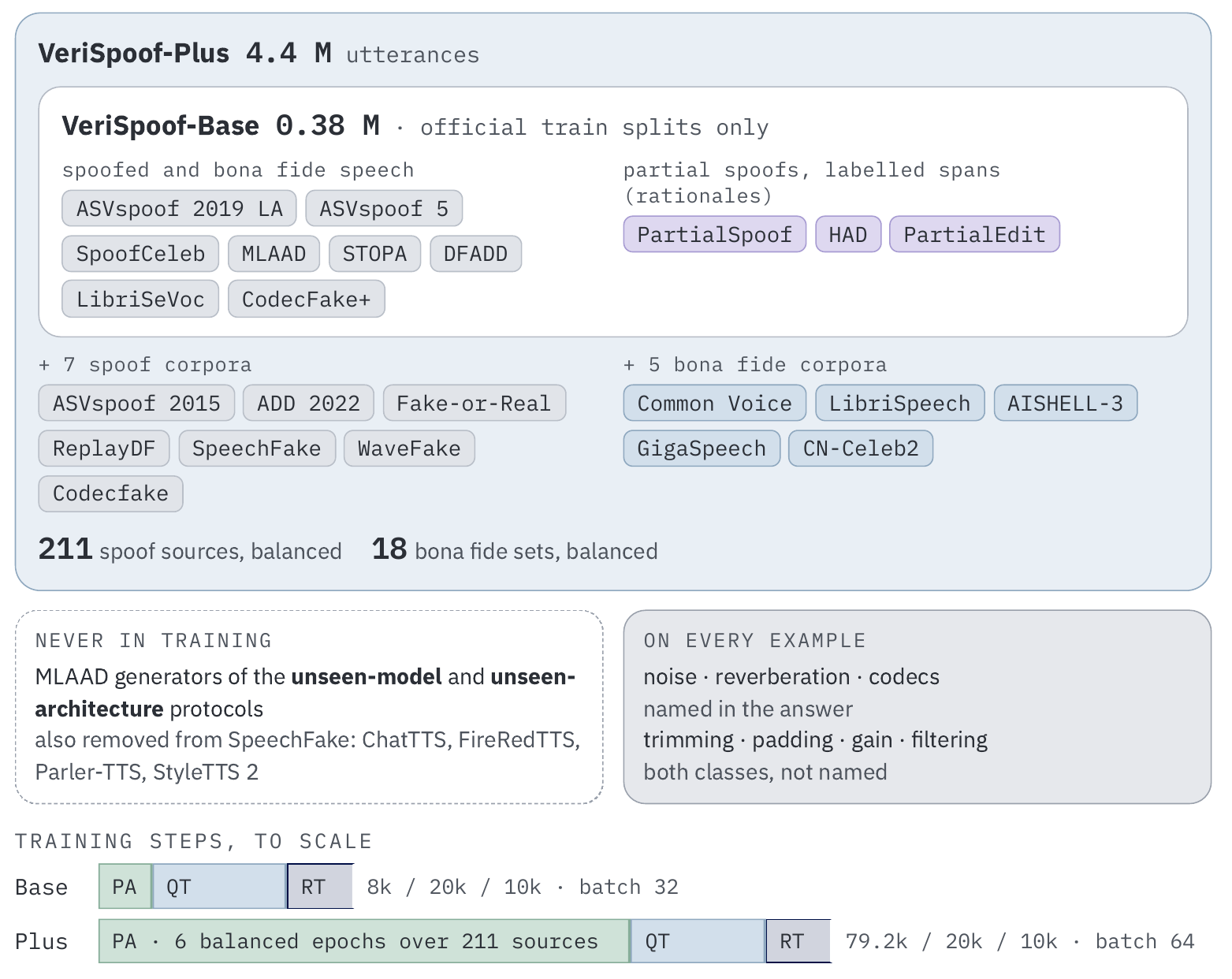}
\caption{\textbf{Training data.} \textit{\name{}-Plus} keeps everything \textit{\name{}-Base} is trained on and adds twelve more
corpora, balanced over 211 spoof sources and 18 bona fide datasets. Grey: corpora with
spoofed speech; purple: partial spoofs whose labelled spans train the rationales; blue: bona fide speech only. Bottom:
training steps of the three stages, perception alignment (PA), forensic question tuning (QT) and grounded rationale
tuning (RT), to scale.}
\label{fig:data}
\end{figure}

Figure~\ref{fig:data} summarizes the training data of both models.

\paragraph{\name{}-Base} Base uses the official training splits of eight detection corpora, ASVspoof 2019 LA
\citep{asvspoof2019}, ASVspoof~5 \citep{asvspoof5}, SpoofCeleb \citep{spoofceleb}, MLAAD \citep{mlaad}, STOPA
\citep{stopa}, DFADD \citep{dfadd}, LibriSeVoc \citep{librisevoc} and CodecFake+ \citep{codecfakeplus}, and the
partially spoofed corpora PartialSpoof \citep{partialspoof}, HAD \citep{had} and PartialEdit \citep{partialedit} for
rationales: about 0.38\,M utterances. No evaluation split of any corpus is used. Base serves every controlled study of
this report, so that its results are not tied to a particular choice of extra data.

\paragraph{\name{}-Plus} Plus keeps the recipe and adds twelve corpora, spoofed speech from ASVspoof 2015 \citep{asvspoof2015}, ADD 2022 train / dev \citep{add2022}, Fake-or-Real
\citep[normalized version;][]{fakeorreal}, ReplayDF \citep{replaydf}, SpeechFake \citep{speechfake}, WaveFake
\citep{wavefake} and Codecfake \citep[codec conditions F01--F06;][]{codecfake}, and bona fide speech from a Common
Voice 15.0 subset \citep{commonvoice}, LibriSpeech \citep{librispeech}, AISHELL-3 \citep{aishell3}, GigaSpeech
\citep{gigaspeech} and CN-Celeb2 \citep{cnceleb}, 4.0\,M utterances more. Its spoofed half is balanced over 211 sources
(one generator of one corpus; MLAAD generators grouped into 82 architecture families) and its bona fide half over 18
datasets, so that the large added corpora do not dominate.

\paragraph{Held-out generators} The MLAAD generators of the unseen-model and unseen-architecture verification
protocols are removed from all training data, including four SpeechFake systems that are among them (ChatTTS,
FireRedTTS, Parler-TTS, StyleTTS~2). Codecfake and CodecFake+ are different corpora: CodecFake+ has a codec
re-synthesis part (``CodecFake+'' in our tables) and a part generated by codec-based speech generators (CoSG), whose
generators are unseen in training.

\paragraph{Augmentation} Training copies carry labelled noise MUSAN \citep{musan}, reverberation and codecs,
including narrowband codecs, drawn anew for every example because the encoder runs inside the training loop. The
recording-condition questions and the \textsc{channel\_status} statements are trained on these labels. Post-processing
(silence trimming or padding, gain with clipping, high- or band-pass filtering) is applied identically to both classes,
without naming it in the answer, which removes channel shortcuts (\S\ref{sec:res-shortcuts}); on the corpora added for
Plus, augmentation is applied without naming it either.

\paragraph{Recipe} Base trains 8k / 20k / 10k steps in the three stages at a global batch of 32 on one 48\,GB GPU, about 14\,GPU-hours
(\S\ref{sec:res-speed}). Plus
runs perception alignment for six balanced epochs (79.2k steps) at a global batch of 64 on two GPUs and keeps the step
counts of the other two stages. Learning rates fall from $10^{-4}$ in perception alignment to $2{\times}10^{-5}$
(LoRA) in rationale tuning. Questions are sampled by task weights; the full settings are in
Table~\ref{tab:app-hparams} and the training data in Figure~\ref{fig:data}.

%% file: sections/evaluation.tex
\section{Evaluation suite}
\label{sec:setup}
\label{sec:evaluation}

\paragraph{Public detection boards} For detection we score every line of the 14 official protocols of
Speech-DF-Arena \citep{dfarena}: ASVspoof 2019 LA (LA19), 2021 LA and DF (LA21, DF21) \citep{asvspoof2021},
ASVspoof~5 (ASV5), In-the-Wild (ITW) \citep{itw}, Fake-or-Real (FoR), Codecfake (Codec), ADD 2022 tracks 1 and 3
(A22T1, A22T3), ADD 2023 rounds 1 and 2 (A23R1, A23R2) \citep{add2023}, DFADD, LibriSeVoc (LSV) and SONAR, 2.34\,M
trials in total. We report the pooled EER, which uses one threshold over all sets and decides the board ranking, and
the mean over sets, and compare with the board's public snapshot of 31 Oct 2026. We also score the 16 sets of the
SpoofRadar v1.0 protocol \citep{spoofradar}, which adds XMAD-Bench \citep{xmadbench} in seven languages and SpoofCeleb,
with its pooled metrics (Table~\ref{tab:app-card}). We did not submit to either leaderboard; our numbers are computed
with the boards' own protocol files.

\paragraph{\name{}-Bench} 

\begin{figure}[t]
\centering
\figcard{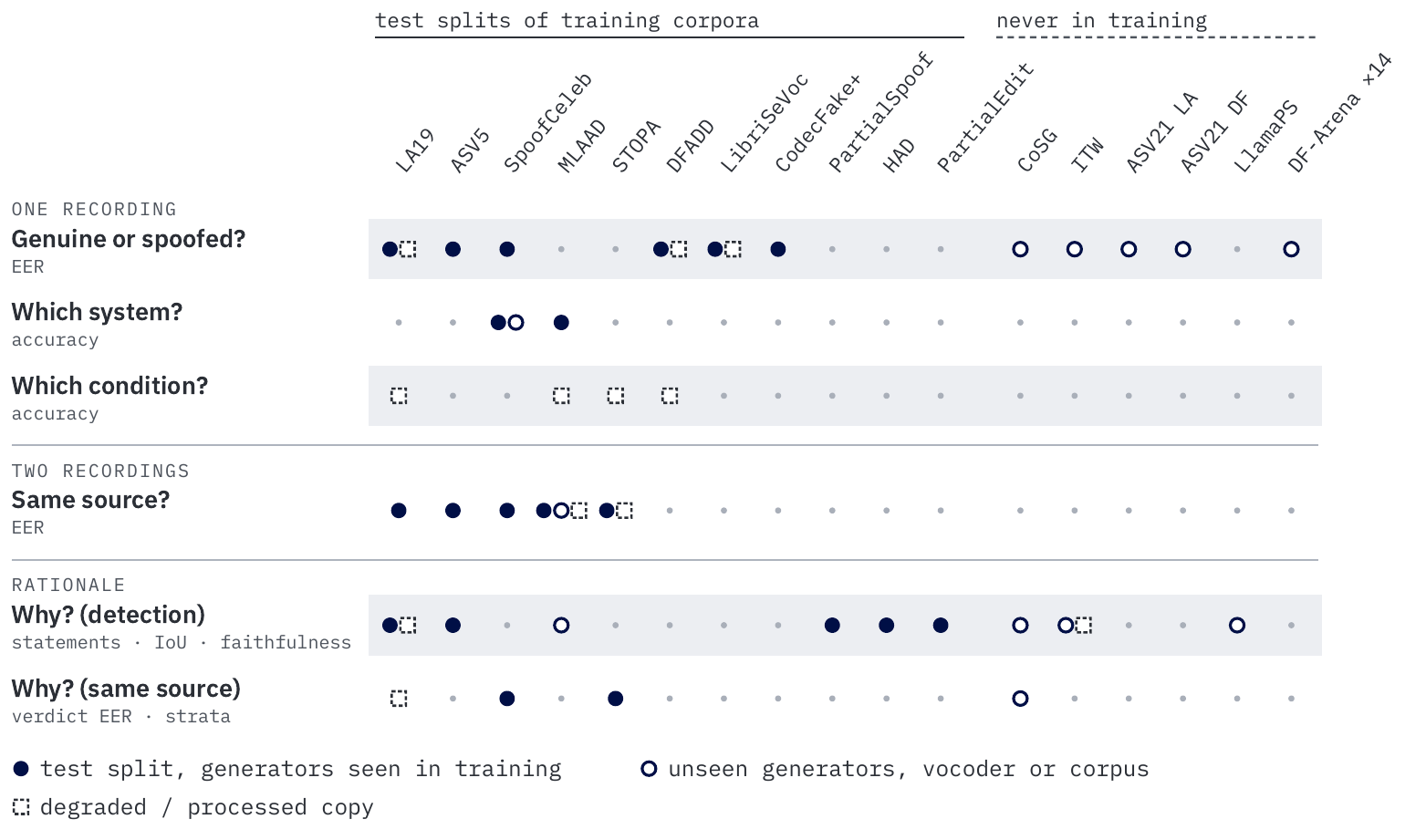}
\caption{\textbf{Questions and their test sets.} Columns on the left are test splits of corpora used in training;
columns on the right are sets never used in training. Filled circle: generators seen in training; open circle:
generators, a vocoder pairing or a whole corpus unseen in training; dashed square: copies with added noise,
reverberation and codecs (aug.), recorded under a logged condition (cond.) or post-processed (+proc). Speech-DF-Arena
adds 14 public sets, scored on their full protocols. Figure~\ref{fig:questions} shows each question with an example
answer.}
\label{fig:tasks}
\end{figure}

The other questions have no public benchmark, so we built one and release \textit{\name{}-Bench}. It has
54 subsets with 177,852 items in six groups (Figure~\ref{fig:tasks}, Figure~\ref{fig:bench}): up to 10,000 items per
closed-answer subset, drawn once and used for every model and checkpoint, so differences between columns of a table
are paired. Source verification on MLAAD is split by how far the generators of a pair are from training:
seen generators, unseen models of a seen architecture, and unseen architectures. Degraded copies come in two kinds:
aug.\ copies apply noise, reverberation and codecs drawn as in training, and cond.\ copies carry one logged condition
per recording, against which channel statements are checked. Rationales are evaluated on 500 utterances per set, and
verification rationales on pairs balanced over the SAME, REVERSAL and DIFF strata. +proc marks copies with
post-processing.

\paragraph{Metrics} Yes--no questions are scored by EER on the score of Eq.~\eqref{eq:score}; questions with more
answers by the accuracy of the most likely option. Observation statements are correct when the stated class matches
the measured one (SNR within 10\,dB, speech level within 5\,dB, bandwidth and clipping classes). Localization is scored
by the IoU of stated and labelled spans, and the per-token head by F1 of the synthesized class on the 0.16\,s grid.
Faithfulness is the share of decisions that follow an edited localization (follow rate) or stay unchanged under an
edited non-evidential statement (stay rate).

\paragraph{Baselines} Detection is compared with the open-weight systems of Speech-DF-Arena and SpoofRadar;
proprietary systems, whose training data are undisclosed, are listed for reference. No open system answers the source
questions, so we train a pair probe on the frozen anti-spoofing encoder DF-Arena-1B with the Base training data, and
prompt six general audio LLMs zero-shot with closed answer options \citep{qwen3omni, minicpmo45, kimiaudio,
audioflamingo3, stepaudio2, mimoaudio}.

\paragraph{Controls} Controlled comparisons change one factor at a time with all else fixed and use the same test
trials. The supervision study (\S\ref{sec:res-supervision}) fixes its decision rule before the results and reports
paired, class-stratified bootstrap intervals.

%% file: sections/results.tex
\section{Results}
\label{sec:results}

\subsection{Detection on public benchmarks}
\label{sec:res-detection}
\label{sec:rq1}

\input{tables/rq1_arena}

\paragraph{Best open-weight detector} On the 14 Speech-DF-Arena sets (Table~\ref{tab:arena}), \textit{\name{}-Plus} has the
lowest mean EER (1.79\%) and pooled EER (2.14\%) of all open-weight systems; the next best, DF-Arena-1B, reaches 5.66\%
and 9.52\%, and every other open-weight system, including four listed only on SpoofRadar, has a mean EER above 5\%.
Inserted into the public snapshot, Plus ranks third of 28 systems by pooled EER, behind two proprietary systems, and
Base fourth (Figure~\ref{fig:glance}a). Base is trained on the official training splits of eight corpora; Fake-or-Real,
Codecfake, ADD and SONAR have no training data in Base, and their EERs are its main gap to the systems above it. The
score is the likelihood of the model's own answer, so answering in sentences costs neither a detection head nor
accuracy.

\paragraph{SpoofRadar and unseen languages} On the 16 sets of the SpoofRadar v1.0 protocol (Table~\ref{tab:app-spoofradar}),
both models have a lower pooled EER (Base 4.23\%, Plus 4.04\%) and higher accuracy and macro-F1 than every open-weight
system on the published leaderboard, the best of which reaches 9.96\%. XMAD-Bench \citep{xmadbench}, seven languages
whose genuine speech comes from M-AILABS \citep{mailabs}, MASC \citep{masc}, VoxPopuli \citep{voxpopuli} and
AISHELL-3, is unseen by both models apart from 1,001 AISHELL-3 recordings, and is detected by Base at 0.24\% EER and by
Plus at 1.33\%.

For Plus, however, the pooled EER (4.04\%) is far above the mean of its set EERs (1.65\%), because one threshold does not fit
every set. At the pooled threshold, 92.9\% of the genuine Russian audiobook recordings of XMAD-Bench (M-AILABS) are
judged spoofed, as are 32.2\% and 15.4\% of the Spanish (M-AILABS) and Arabic (MASC) ones, although within each language the
two classes are separated at $\leq$1.95\% EER. Base misclassifies at most 3.1\% of the genuine recordings of any
language. The data added for Plus thus shift its scores on unseen genuine speech without making it less separable, a
calibration cost that pooled metrics expose and per-set EER hides. A deployment on a new language or domain should
therefore set its threshold on genuine speech of that domain.

\input{tables/app_spoofradar}

\paragraph{What more data changes} From Base to Plus, 9 of the 14 common sets improve and 3 get worse. Gains are
largest where Plus added training data of the same corpus: Codecfake 4.97 $\to$ 0.12 (Codecfake train),
Fake-or-Real 5.39 $\to$ 0.09 (for-norm train), ADD 2023 rounds 1 and 2 5.01 / 10.61 $\to$ 0.50 / 2.41 (ADD 2022
train). Losses are on ASVspoof~5 (2.59 $\to$ 3.40), ASVspoof 2021 DF (0.97 $\to$ 1.24) and In-the-Wild (1.20 $\to$
1.90): ASVspoof~5 shrinks from half of the training data to 8 of 211 balanced sources. The data mixture decides what
the model is best at.

\paragraph{Degraded copies and the cost of rationales} On the \textit{\name{}-Bench} (Table~\ref{tab:rq1-det}), Plus is better
on 12 of 16 rows and, as on the public sets, worse on ASVspoof~5 (2.23 $\to$ 3.41) and In-the-Wild (1.25 $\to$ 1.83);
its largest gains are on degraded copies (LibriSeVoc aug.\ 10.31 $\to$ 7.72, LA19 cond.\ 5.00 $\to$ 3.87), where more
varied training channels help. Between question tuning and rationale tuning, detection changes by less than 0.3 points
on most sets, so adding rationales does not cost one-step detection; the exceptions are ASVspoof 2021 LA (3.40 $\to$
5.17) and CodecFake+ (2.24 $\to$ 3.95), neither of which supplies rationale data.

\input{tables/rq1_detection}

\subsection{Source verification and attribution}
\label{sec:res-source}

\input{tables/rq1_source}

\paragraph{The same weights verify sources} For two recordings from generators unseen in training
(Table~\ref{tab:rq1-src}), \textit{\name{}-Plus} verifies a shared source at 6.17\% EER when the generators are new models of
known architectures and at 15.00\% when the architectures are new, which suggests that source cues are tied to
architectures. It names seen generators (96.8\%) and recording conditions (81.3\%). Degraded copies remain hard
(23--30\% EER), as noise and codecs mask the detail that separates generators. The pair probe on the frozen
DF-Arena-1B encoder reaches only 24.8--29.0\% EER on the MLAAD sets; Base is better on every MLAAD, SpoofCeleb and
degraded set and worse on STOPA (20.88 vs.\ 17.9) and ASVspoof~5 (23.27 vs.\ 22.2), and Plus is better on every set.

\paragraph{General audio LLMs cannot answer these questions} Prompted with the same closed options on 1,000 items per
set (Table~\ref{tab:app-zeroshot}), six general audio LLMs answer the same-source and recording-condition questions
near chance (48.4--53.3 and 16.4--22.6\% balanced accuracy on average). On authenticity they reach 52.1--64.7\%; the
best, Qwen3-Omni, reaches 76.0--90.2\% on the ASVspoof sets but stays near chance on SpoofCeleb, LibriSeVoc and both
CodecFake+ sets. On the same items Base and Plus reach 95.9 and 96.5\% on authenticity, 77.9 and 82.8\% on same source,
and 76.9 and 83.0\% on recording condition.

\input{tables/app_zeroshot}

\paragraph{More generators sharpen the fingerprint} Plus improves source verification on 10 of 11 sets
(Table~\ref{tab:app-src}), by 6--7 points on degraded copies (unseen models 13.80 $\to$ 6.17\%): seeing more generators
sharpens the notion of a source fingerprint. It loses on SpoofCeleb (11.25 $\to$ 12.20) and on naming the acoustic
model under an unseen vocoder (84.7 $\to$ 69.8\%), the two tasks that rest on SpoofCeleb's labels, whose share of the
training data is far smaller in Plus. Rationale tuning keeps the source answers of Base within 1.3 EER points and 2.4
accuracy points of question tuning, so they survive the last stage through replay.

\paragraph{Composition depends on the LLM} Naming the acoustic model of SpoofCeleb systems A17 and A18 requires
composing an acoustic model seen in training with a vocoder never seen with it (Table~\ref{tab:comp}). Linear probes
recover the acoustic model from encoder states for every LLM run (0.84--0.90) and less well from pooled audio tokens
(0.64--0.75), yet the answers range from chance (Qwen2.5-1.5B, 0.52) to 0.86 (MiniCPM5-2B, above its token probe). The
information reaches the LLM in all runs; whether it is composed into an answer depends on the LLM.

\input{tables/app_source_1b}
\input{tables/analysis_composition}

\subsection{What determines what the model perceives}
\label{sec:res-perception}
\label{sec:rq2}

\begin{figure}[t]
\centering
\figcard{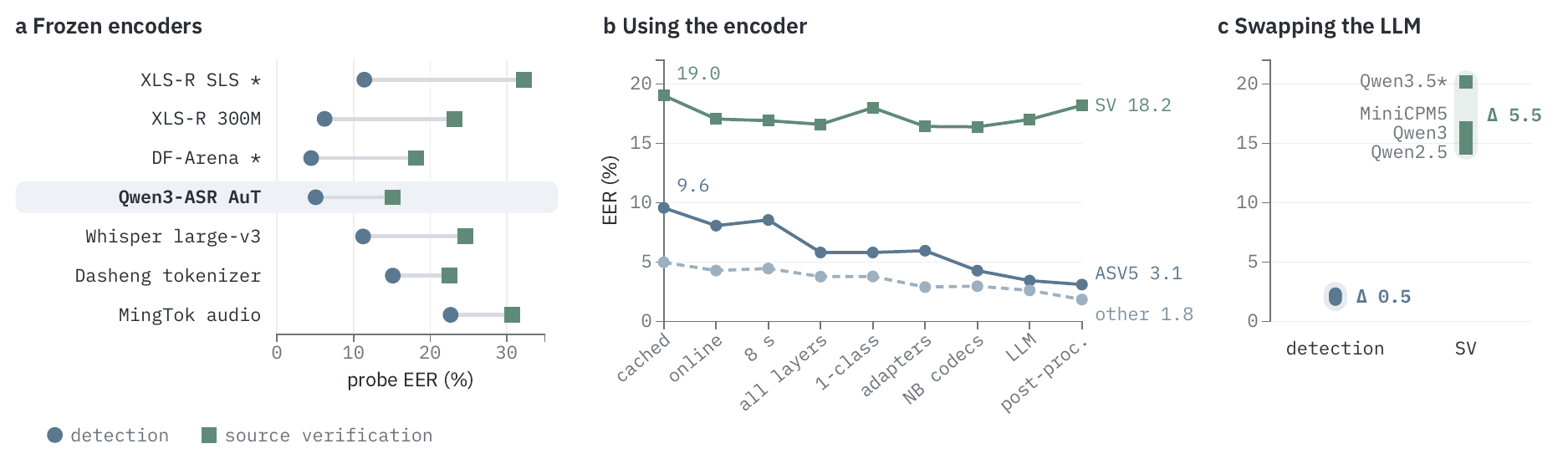}
\caption{\textbf{Components.} (a) Probes on frozen features of seven encoders (best single layer; same items for every
encoder; * trained for anti-spoofing). (b) One factor changed at a time in how the encoder is used, LLM frozen:
detection EER on ASVspoof~5 falls threefold, source verification (SV) stays flat. (c) Only the LLM changed: detection
moves by 0.5 points, SV by 5.5; * hybrid linear attention. (a) uses an earlier protocol; Table~\ref{tab:app-probe}
repeats it for two encoders on the current test sets.}
\label{fig:rq2}
\end{figure}

An ALM perceives through its encoder, tokenizer and LLM; we ask which of these limits detection and which limits
source comparison.

\paragraph{The speech-recognition encoder keeps the evidence} Probes on frozen features (Figure~\ref{fig:rq2}a) show
that Qwen3-ASR gives the best source verification of seven encoders and detection close to the best anti-spoofing
encoder, while audio tokenizers lose much of both. Repeated under the current protocol against the strongest
anti-spoofing encoder, DF-Arena-1B (Table~\ref{tab:app-probe}), Qwen3-ASR with a learned layer mix verifies sources
better (mean EER 24.7 vs.\ 26.4\%, better on 11 of 15 pair sets), and DF-Arena-1B detects better on average (13.3 vs.\
14.7\%), most clearly on ASVspoof 2021 LA, CoSG and In-the-Wild; the best single layer of Qwen3-ASR is layer 12 of 25
for both tasks. The view that speech-recognition encoders discard the artifacts of synthesis \citep{alethia} does not
hold for this encoder.

\paragraph{Detection depends on how features are used; source identity does not} With the encoder fixed, a ladder of
changes in how its features are used (Table~\ref{tab:rq2}a, Figure~\ref{fig:rq2}b) lowers clean detection EER from
4.97 to 1.84\% and ASVspoof~5 from 9.56 to 3.10\%. The largest single steps are fusing all 25 encoder layers
(ASVspoof~5 8.54 $\to$ 5.80), encoder DoRA (degraded detection 8.39 $\to$ 6.33) and narrowband codecs in augmentation
(ASVspoof~5 5.95 $\to$ 4.27). Clean source verification stays between 16.4 and 19.0\% throughout. Local, fragile
synthesis artifacts gain from varied channels and multi-layer detail, whereas source identity is a global property the
features already expose. The last step, post-processing applied to both classes, is the best for detection but raises
source-verification EER (17.00 $\to$ 18.19 clean); we keep it because it removes the channel shortcuts of
\S\ref{sec:res-shortcuts}.

\paragraph{Comparison depends on the LLM} With the representation fixed, swapping the LLM (Table~\ref{tab:rq2}b,
Figure~\ref{fig:rq2}c) moves mean detection EER by 0.5 points but clean source verification by 5.5. The hybrid
linear-attention LLM, Qwen3.5-4B \citep{qwen35}, is the best detector and the worst comparer, as relating two token
sequences needs recall that such attention lacks. Qwen2.5-1.5B \citep{qwen25} has the best clean source verification
(14.61\%) but states spans poorly (IoU 0.36) and names the acoustic model less often (0.77); MiniCPM5-2B is best on
degraded pairs, localization and the acoustic model, and is used for both released models.

\input{tables/rq2_main}

\subsection{Channel shortcuts}
\label{sec:res-shortcuts}
Without post-processing in training, the model reads channel properties as evidence (Table~\ref{tab:rq2-short}): a
2\,kHz high-pass makes 11.6--26.8\% of spoofs pass as genuine, and 0.5\,s of digital silence or +10\,dB gain with
clipping makes 12.4 and 15.6\% of genuine ASVspoof~5 speech look spoofed. Applying the same post-processing to both
classes reduces these to 1.7--8.4\%. Labelling the post-processing in the answer gives no consistent further gain
(1.9--8.9\%), so what removes the shortcut is that both classes share the channel, not that the model is told about
it; we keep the unlabelled form. Gain with clipping remains the largest error (8.4\%).

\input{tables/rq2_shortcut}

\subsection{Does knowing the source help detection?}
\label{sec:res-supervision}
\label{sec:rq3}

\begin{figure}[t]
\centering
\figcard{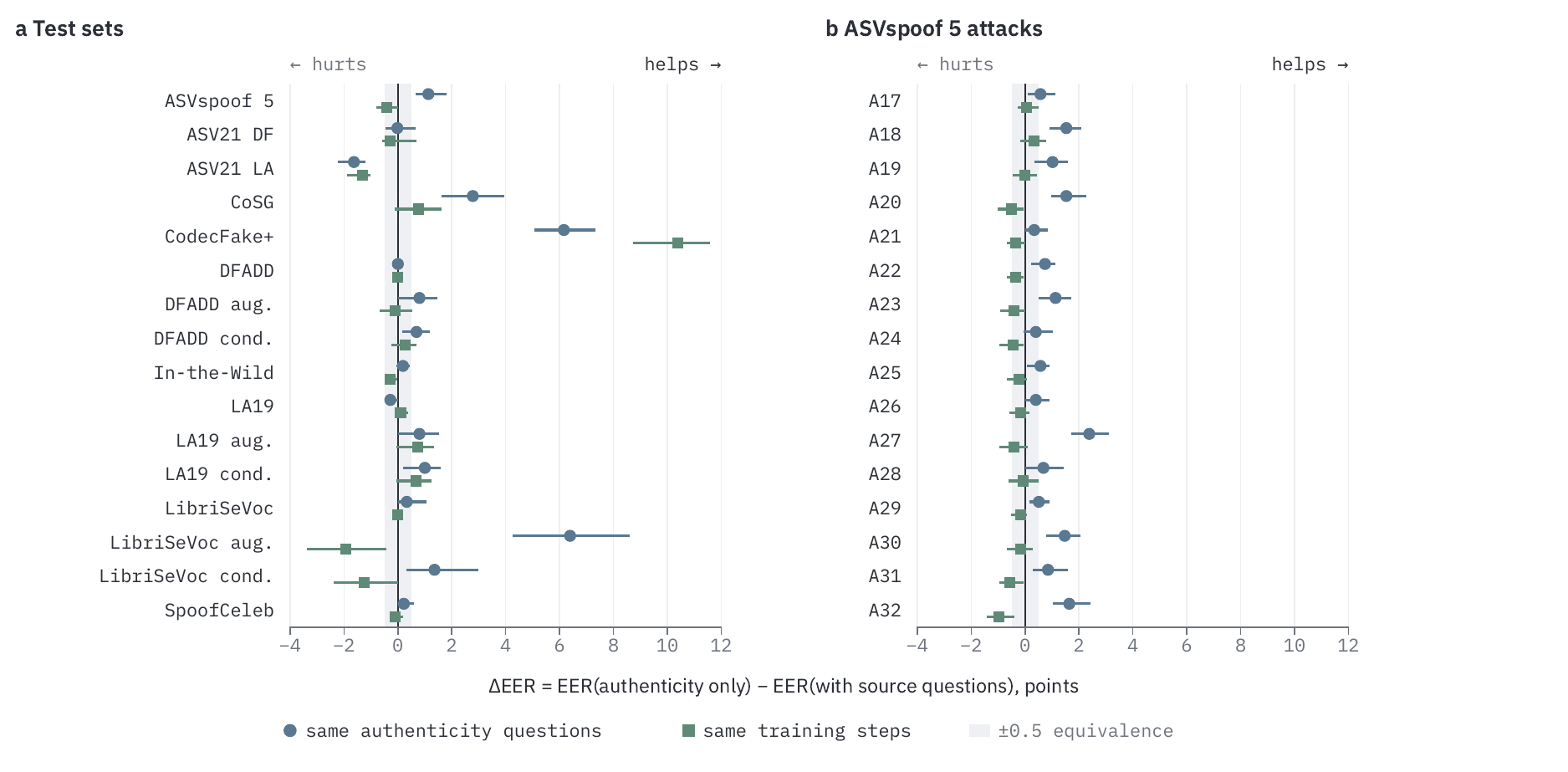}
\caption{\textbf{Source supervision} per test set (a) and per ASVspoof~5 attack (b; A17--A32 are the attack
identifiers of the ASVspoof~5 evaluation set): EER without source questions minus EER with them (95\% bootstrap
intervals; $>0$: source questions help). Circles: same number of authenticity questions (B); squares: same number of
training steps (B$'$); shaded: the $\pm0.5$ equivalence margin.}
\label{fig:rq3}
\label{fig:app-attacks}
\end{figure}

A spoof labelled with its generator says more than ``fake'', but earlier evidence that source questions help detection
did not match authenticity labels or steps \citep{holiantispoof}. We compare the multi-task model A (Base after
question tuning) with two authenticity-only models: B sees as many authenticity questions as A, and B$'$ trains as many
steps and so sees five times more. For each test set and ASVspoof~5 attack we compute $D = \mathrm{EER}_{B} -
\mathrm{EER}_{A}$ with a paired bootstrap interval and apply a rule fixed before the results: helps (interval above
zero), hurts (below), equivalent (inside $\pm0.5$) or inconclusive.

\paragraph{Per label it helps; at equal compute it mostly breaks even} Against B (Figure~\ref{fig:rq3}, circles;
Table~\ref{tab:rq3}), source supervision helps on 8 of 16 sets and 12 of 16 ASVspoof~5 attacks and hurts on no
attack, most on codec-based generators (CoSG $+2.8$, CodecFake+ $+6.2$) and degraded copies. Against B$'$ (squares),
most differences become equivalent or inconclusive: source labels act mainly by making each example more informative.
More binary examples help where the fakes are whole-utterance and the variation is in the channel (In-the-Wild
$-0.28$, LibriSeVoc aug.\ $-1.94$ against B$'$).

\paragraph{For neural-codec fakes it is necessary} With five times more authenticity questions, B$'$ still has 12.6\%
EER on CodecFake+ against 2.2\% for A ($D = 10.4$ [8.7, 11.6]). Codec resynthesis resembles lossy compression of real
speech, so a binary label gives no reason to separate them, whereas asking which codec produced a recording does; this
complements the finding that authenticity-only detectors encode generator identity implicitly \citep{doss}.

\paragraph{An open cost on ASVspoof 2021 LA} ASVspoof 2021 LA is the one set where source supervision hurts ($-1.6$
against B, $-1.3$ against B$'$). The loss sits in three attacks unseen in training (A10, A11 and A18, attack identifiers
of ASVspoof 2019 LA that ASVspoof 2021 LA reuses) and grows from question tuning to rationale tuning (3.40 $\to$
5.17\%, Table~\ref{tab:rq1-det}); it is not a calibration shift and not tied to one codec. DFADD and LibriSeVoc are at
0\% EER for every model and cannot separate the conditions.

\input{tables/rq3}

\subsection{Every statement can be checked}
\label{sec:res-rationale}

\input{tables/app_grounded}

\paragraph{Most statements are right} Speech level is stated correctly for at least 95\% of recordings and SNR for
88--99\% (Table~\ref{tab:app-grounded}); clipping is weakest on ASVspoof~5 (0.73), and the bandwidth class is the least
reliable statement (0.64--0.91). Probes show that finer bandwidth classes are not recoverable from the audio tokens
(38\% for 0.5\,kHz bins), so coarse classes are the resolution the model can support. A wrong statement fails its
check, a signal that free-text explanations lack: in Figure~\ref{fig:teaser} the bandwidth statement fails while the
span and the decision hold.

\paragraph{Spans: the frames carry the evidence, writing them is the bottleneck} Stated spans reach IoU 0.58--0.61 on
HAD and PartialEdit, where a recording has one or few edited segments, but only 0.25 and 0.31 on PartialSpoof and
LlamaPartialSpoof \citep{llamapartialspoof}, which have many short ones. The per-token head on the same hidden states
reaches F1 0.74 and 0.82 there, so the frames are separable and writing many short spans as text is the bottleneck.
Span IoU also varies with the LLM (0.25--0.64 on HAD), whereas the per-token head varies little (0.89--0.91;
Table~\ref{tab:rq4-loc}): the LLM decides how precisely spans are written, not whether the frames carry the evidence.

\input{tables/rq4_loc}

\subsection{Reasoning versus the one-step answer}
\label{sec:res-reason}

\begin{figure}[t]
\centering
\figcard{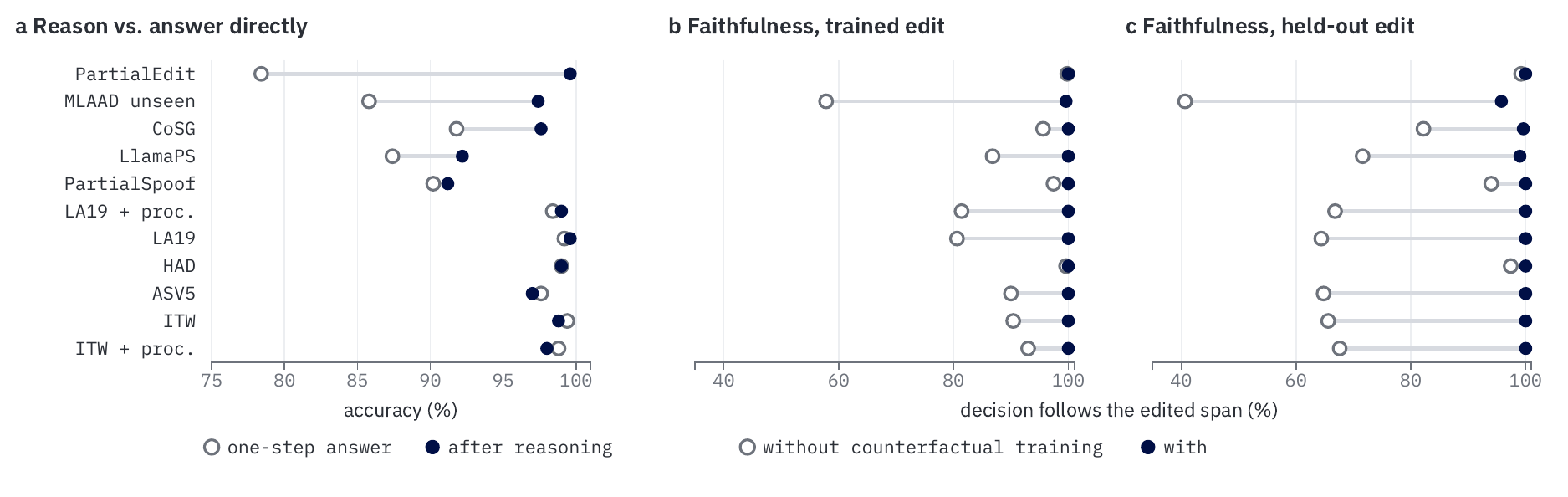}
\caption{\textbf{Reasoning and faithfulness.} (a) Accuracy of the one-step answer and of the decision after reasoning on
the same 500 recordings per set (\textit{\name{}-Base}). (b, c) Share of decisions that follow a localization replaced by its
opposite, without and with counterfactual training, for the edit form used in training (b) and a paraphrased form
never seen in training (c).}
\label{fig:rq4}
\end{figure}

\paragraph{Reasoning catches what the one-step answer misses} The decision after reasoning agrees with the one-step
answer on 97.6--99.6\% of whole-utterance spoofs of known corpora, and where they disagree the reasoning is usually
right (Figure~\ref{fig:rq4}a, Table~\ref{tab:app-agree}): 99.6 vs.\ 78.4\% on PartialEdit, 97.4 vs.\ 85.8\% on unseen
MLAAD generators. The one-step answer summarizes the whole recording, whereas writing the localization first makes the
model inspect it frame by frame. Unlike the chain-of-thought of ThinkOmni, which lowered accuracy \citep{thinkomni},
our intermediate steps are perception targets with checks.

For Plus the one-step answer is near chance on HAD (48.4\%) and 80.0\% on PartialSpoof, while its reasoning decision
reaches 97.6 and 89.2\%: the data added for Plus shift the one-step answer on partial fakes, but writing the
localization first still finds the edited span. On whole-utterance sets the one-step answer of Plus catches up (MLAAD
99.4\%), and the two agree on 98.4--99.6\%. Reasoning is therefore most useful where the training data of the
one-step answer do not cover the fake. In practice, the one-step score suits screening at scale, and a rationale suits
the recordings an analyst has to decide on.

\input{tables/app_agree}

\paragraph{The order of statements matters} Comparing attributes before the verdict led verification rationales to
answer ``same source'' whenever attributes matched, failing on REVERSAL pairs (accuracy 0.31) although the one-step
answer was good (Table~\ref{tab:rq4-svr}); in two rationale-tuning runs that differ only in field order, stating the
fingerprint verdict first raised this to 0.67. Qwen2.5-1.5B reaches 0.76 even with attributes first, so the failure of
the attribute-first order depends on the LLM. The final Base, which states the fingerprint verdict first, falls back to
0.42, together with its one-step EER on the same pairs (28.45 vs.\ 18.35\%), which points to its perception of these
pairs rather than the format; Plus reaches 0.75. The statement that carries the decision must precede those that could
mislead it.

\input{tables/rq4_svr}

\subsection{The decision follows the stated evidence}
\label{sec:res-faith}
Without counterfactual training, the model told that no segment was synthesized still answered ``spoof'' for 42\% of
unseen MLAAD recordings (Figure~\ref{fig:rq4}b); with it, the decision follows the edit on every set. This carries
over to edits never seen in training (Figure~\ref{fig:rq4}c): a paraphrased opposite localization is followed in
99.5\% of cases (73.8\% without counterfactual training), an inserted span in 100\%, and the decision stays unchanged
in 100\% when channel or observation statements are contradicted. The decision thus depends on the localization, which
is the statement that should carry it, and not on the statements declared non-evidential. The model is never trained on a
false statement about the audio, and the statement accuracies of Table~\ref{tab:app-grounded} are those of the model
trained with counterfactual examples.

\subsection{Compute: training and running on small GPUs}
\label{sec:res-speed}
\paragraph{Only 1\% of the parameters are trained} \textit{\name{}} trains 29.7\,M of its 2.85\,B parameters: the audio
tokenizer (17.85\,M, all stages), the encoder DoRA adapters (3.24\,M, perception alignment), the LLM LoRA adapters
(8.60\,M, question and rationale tuning) and the per-token head (6\,k); a one-class head of 0.53\,M is used in training
only. The encoder and the LLM stay frozen in bf16 and hold no gradients or optimizer state, so the training state of the
adapters (fp32 copy, gradients and Adam moments) takes about 0.5\,GB. Memory is set by the frozen weights, about
5.3\,GiB in bf16, and by activations.

\paragraph{Training} Table~\ref{tab:memory} gives the peak memory logged in our runs on one 48\,GB L40S. The
released Base was trained without gradient checkpointing, which peaks at 21--45\,GiB and takes about 14\,GPU-hours for
the three stages (logged throughput 50, 24 and 18 examples per second). With gradient checkpointing, runs of the same
model and stages at the same micro-batch of 32 peak at 11--17\,GiB, so the full recipe, global batch included, fits a
single 24\,GB GPU such as an RTX 3090 or 4090, at a lower speed. Activations grow with the micro-batch, about
0.17--0.34\,GiB per example; we estimate that a micro-batch of 8 with gradient checkpointing, accumulated to the same
global batch, needs 7--9\,GiB per stage, within a 12\,GB GPU, and a micro-batch of 1--2 approaches the 5.3\,GiB of
frozen weights. Offloading the frozen weights to CPU memory, for example with DeepSpeed ZeRO-Offload, would lower the
floor further at the cost of speed. These settings change the micro-batch only, not the recipe; we have not trained at
them. The code runs in bf16, which needs an Ampere or newer GPU (RTX 30-series, A10, L4 or later); GPUs without bf16
such as the T4 or V100 would need fp16 or fp32, which we have not tested.

\begin{table}[h]
\centering
\small
\caption{Peak GPU memory in training (GiB), global batch 32 throughout. Logged on one L40S: the runs of the released
\textit{\name{}-Base} (no gradient checkpointing; micro-batch 32 in perception alignment, 8 with accumulation in the other
stages), and runs of the same model and stages with gradient checkpointing at a micro-batch of 32. Estimated: micro-batch 8 with checkpointing, from the frozen weights (5.3\,GiB), the adapter state
(0.5\,GiB) and the logged activation memory per example.}
\label{tab:memory}
\begin{tabular}{lccc}
\toprule
Stage & \multicolumn{2}{c}{Logged} & Estimated \\
\cmidrule(lr){2-3}
 & released Base, no ckpt & micro-batch 32, ckpt & micro-batch 8, ckpt \\
\midrule
Perception alignment & 45.3 & 16.8 & $\approx$\,9 \\
Forensic question tuning & 21.4 & 11.1 & $\approx$\,7 \\
Grounded rationale tuning & 24.6 & 14.4 & $\approx$\,8 \\
\bottomrule
\end{tabular}
\end{table}

\paragraph{Inference} All 2.85\,B parameters are loaded in bf16, about 5.3\,GiB. Scoring one 8\,s recording and
generating a rationale of a few hundred tokens add well under 1\,GiB of activations and cache, so we estimate 6--7\,GiB
for single recordings: an 8\,GB GPU such as an RTX 3060 Ti or 4060 is enough, and about 12\,GB covers batched detection
(16 recordings per batch by default). Quantizing the LLM to 8 or 4 bits would lower this further, which we have not
evaluated. Rationales are the expensive output, as they are generated;
the vLLM backend generates them from the same prompt embeddings, with the audio front end in PyTorch
(Table~\ref{tab:speed}). Decisions are identical to the reference implementation on 64 of 64 recordings and 32 of 32
pairs; the wording of borderline measurements can differ.

\begin{table}[h]
\centering
\small
\caption{Rationale generation time of \textit{\name{}-Base} on one A100 GPU.}
\label{tab:speed}
\begin{tabular}{lcc}
\toprule
Rationale generation & per recording & speed-up \\
\midrule
PyTorch, one recording at a time & 2.24\,s & 1$\times$ \\
vLLM, one recording at a time & 0.50\,s & 4.5$\times$ \\
vLLM, batch of 64 & 0.05\,s & 46$\times$ \\
\bottomrule
\end{tabular}
\end{table}

%% file: tables/rq1_arena.tex
\begin{table*}[t]
\centering
\small
\caption{Detection on the 14 Speech-DF-Arena sets (EER, \%; best per column in bold). Speech-DF-Arena snapshot of 31 Oct 2026; Teffic-Audio from its paper; $^\ddagger$ from SpoofRadar, whose pooled EER covers other sets. Pooled: one threshold over all 14 sets. Systems: Modulate \citep{modulate_velma}, Teffic-Audio \citep{teffic}, Hiya \citep{hiya}, DF-Arena \citep{dfarena}, GRL \citep{grl}, XLS-R + SLS \citep{xlsrsls}, XLS-R + MultiConv \citep{xlsr_multiconv}, XLSR-Mamba \citep{xlsrmamba}, XLSR-MamBo \citep{xlsrmambo}, P3 \citep{p3}, Nes2Net-X \citep{nes2net}, AASIST \citep{aasist}.}
\label{tab:arena}
\begin{adjustbox}{max width=\textwidth}
\begin{tabular}{lcccccccccccccccc}
\toprule
System & ITW & LA19 & LA21 & DF21 & ASV5 & FoR & Codec & A22T1 & A22T3 & A23R1 & A23R2 & DFADD & LSV & SONAR & Mean & Pooled \\
\midrule
\multicolumn{17}{@{}l}{\textit{Proprietary}} \\
Modulate VELMA-2 & 1.27 & 0.30 & 1.33 & 0.33 & \textbf{0.38} & 0.13 & 1.54 & \textbf{5.06} & 1.17 & 1.04 & \textbf{1.74} & \textbf{0.00} & 0.27 & 0.89 & \textbf{1.10} & 1.59 \\
Teffic-Audio & 1.32 & 0.24 & 2.04 & \textbf{0.26} & 1.41 & 0.97 & 0.75 & 7.03 & \textbf{0.84} & \textbf{0.36} & 1.84 & \textbf{0.00} & \textbf{0.00} & \textbf{0.25} & 1.24 & \textbf{1.45} \\
Hiya Authenticity v1 & \textbf{0.67} & 0.30 & 1.01 & 1.32 & 0.79 & \textbf{0.00} & 5.73 & 12.10 & 1.19 & 1.98 & 4.01 & 0.02 & \textbf{0.00} & 0.48 & 2.11 & 2.32 \\
\midrule
\multicolumn{17}{@{}l}{\textit{Open weights}} \\
DF-Arena-1B & 0.91 & 1.14 & 4.66 & 1.75 & 17.25 & 2.92 & 8.37 & 22.21 & 2.20 & 5.08 & 11.54 & \textbf{0.00} & 0.15 & 1.09 & 5.66 & 9.52 \\
DF-Arena-500M & 1.76 & 1.09 & 4.23 & 3.30 & 12.39 & 2.30 & 6.36 & 23.98 & 2.77 & 7.47 & 12.30 & \textbf{0.00} & 0.12 & 1.90 & 5.71 & 10.88 \\
GRL$^\ddagger$ & 3.07 & 4.46 & 14.07 & 5.14 & 13.54 & 13.79 & 19.91 & 22.58 & 5.97 & 18.44 & 31.78 & 0.27 & 0.84 & 1.09 & 11.07 & -- \\
XLS-R + SLS & 7.46 & 0.23 & 2.87 & 1.91 & 18.76 & 5.08 & 33.44 & 33.95 & 15.74 & 19.37 & 21.10 & 7.54 & 1.97 & 24.72 & 13.87 & 16.08 \\
XLS-R + MultiConv$^\ddagger$ & 5.53 & 0.41 & 4.88 & 1.92 & 16.16 & 5.91 & 37.43 & 31.76 & 14.56 & 21.39 & 20.45 & 7.38 & 2.16 & 28.44 & 14.17 & -- \\
XLSR-Mamba & 6.71 & 0.42 & 0.93 & 1.88 & 14.40 & 6.71 & 35.27 & 34.23 & 19.37 & 21.85 & 20.15 & 10.70 & 2.23 & 24.26 & 14.22 & 20.59 \\
XLSR-MamBo$^\ddagger$ & 7.29 & 0.20 & \textbf{0.88} & 4.40 & 14.14 & 5.65 & 37.36 & 27.54 & 9.28 & 24.62 & 20.93 & 19.72 & 5.64 & 33.94 & 15.11 & -- \\
P3$^\ddagger$ & 5.86 & 1.73 & 15.48 & 5.87 & 20.03 & 19.92 & 26.89 & 39.73 & 14.30 & 25.98 & 29.03 & 4.90 & 0.65 & 7.58 & 15.57 & -- \\
Nes2Net-X & 7.75 & \textbf{0.12} & 2.17 & 1.49 & 22.06 & 6.32 & 39.34 & 34.47 & 26.56 & 21.14 & 18.45 & 11.15 & 2.88 & 31.54 & 16.10 & 17.37 \\
AASIST & 43.01 & 0.83 & 11.46 & 21.07 & 35.53 & 21.64 & 51.06 & 47.92 & 33.19 & 47.76 & 32.47 & 41.86 & 38.02 & 57.47 & 34.52 & 32.77 \\
\midrule
\multicolumn{17}{@{}l}{\textit{Ours}} \\
\rowcolor{oursbg}\name{}-Base & 1.20 & 0.49 & 5.48 & 0.97 & 2.59 & 5.39 & 4.97 & 14.05 & 3.87 & 5.01 & 10.61 & \textbf{0.00} & 0.02 & 5.19 & 4.28 & 4.43 \\
\rowcolor{oursbg}\name{}-Plus & 1.90 & 0.22 & 3.44 & 1.24 & 3.40 & 0.09 & \textbf{0.12} & 8.65 & 1.18 & 0.50 & 2.41 & \textbf{0.00} & 0.02 & 1.92 & 1.79 & 2.14 \\
\bottomrule
\end{tabular}
\end{adjustbox}
\end{table*}

%% file: tables/app_spoofradar.tex
\begin{table}[!htbp]
\centering
\small
\caption{SpoofRadar v1.0 protocol (16 sets: the 14 of Table~\ref{tab:arena} with LibriSeVoc restricted to its test part, plus XMAD-Bench and SpoofCeleb). Pooled EER over all trials (\%), and accuracy and macro-F1 (\%) at the pooled-EER threshold, as defined by SpoofRadar; Avg.: mean of the 16 set EERs. Other rows: the published leaderboard (Leaderboard v1.0); ours computed with the same protocol files and labels from the source corpora. With one threshold and two classes, micro-F1 equals accuracy. Systems: Teffic-Audio \citep{teffic}; Teffic Labs' closed-source systems built on WavLM-L \citep{wavlm}, Whisper \citep{whisper} and AASIST; DF Arena 500M \citep{dfarena}, GRL \citep{grl}, XLSR-SLS \citep{xlsrsls}, XLSR-MamBo \citep{xlsrmambo}, XLSR + MultiConv \citep{xlsr_multiconv}, NaCL \citep{nacl}, TFCL \citep{tfcl}, Nes2Net-X \citep{nes2net}, XLSR-Mamba \citep{xlsrmamba}, P3 \citep{p3}, W2V2-AASIST \citep{w2v2aasist}, RawNet2 \citep{rawnet2}, AASIST \citep{aasist}.}
\label{tab:app-spoofradar}
\begin{adjustbox}{max width=\linewidth}
\begin{tabular}{lcccccc}
\toprule
System & Pooled EER & ACC & Macro-F1 & Avg.\ EER & XMAD & SpoofCeleb \\
\midrule
\multicolumn{7}{@{}l}{\textit{Commercial}} \\
Teffic-Audio & \textbf{1.37} & \textbf{98.63} & \textbf{98.12} & \textbf{1.09} & \textbf{0.02} & 0.07 \\
\midrule
\multicolumn{7}{@{}l}{\textit{Closed source}} \\
WavLM-L (Teffic) & 3.65 & 96.35 & 95.06 & 2.88 & 0.19 & 0.48 \\
Whisper (Teffic) & 5.75 & 94.25 & 92.33 & 3.94 & 0.27 & 0.13 \\
AASIST (Teffic) & 11.52 & 88.48 & 85.20 & 7.55 & 2.28 & 1.55 \\
\midrule
\multicolumn{7}{@{}l}{\textit{Open weights}} \\
DF Arena 500M V1 & 9.96 & 90.04 & 87.08 & 5.19 & 0.96 & 0.08 \\
GRL & 13.25 & 86.75 & 83.17 & 9.90 & 1.64 & 1.77 \\
XLSR-SLS & 15.97 & 84.28 & 80.25 & 14.07 & 9.13 & 24.52 \\
XLSR-MamBo & 16.33 & 83.67 & 79.64 & 15.17 & 13.58 & 17.53 \\
XLSR + MultiConv & 17.05 & 82.95 & 78.83 & 14.25 & 9.36 & 20.22 \\
NaCL & 17.22 & 82.78 & 78.64 & 16.83 & 14.11 & 22.81 \\
TFCL & 17.84 & 82.16 & 77.95 & 17.15 & 16.31 & 25.29 \\
Nes2Net-X & 17.97 & 82.03 & 77.81 & 18.23 & 10.51 & 54.97 \\
XLSR-Mamba & 19.47 & 80.53 & 76.16 & 18.91 & 15.84 & 27.58 \\
P3 & 19.49 & 80.51 & 76.13 & 15.18 & 9.05 & 15.92 \\
W2V2-AASIST & 20.25 & 79.76 & 75.31 & 18.94 & 21.50 & 27.16 \\
RawNet2 & 33.36 & 66.64 & 61.85 & 36.77 & 29.81 & 57.10 \\
AASIST & 33.66 & 66.34 & 61.56 & 36.08 & 48.37 & 58.26 \\
\midrule
\multicolumn{7}{@{}l}{\textit{Ours}} \\
\rowcolor{oursbg}\name{}-Base & 4.23 & 95.77 & 94.29 & 3.76 & 0.24 & 0.20 \\
\rowcolor{oursbg}\name{}-Plus & 4.04 & 95.96 & 94.54 & 1.65 & 1.33 & \textbf{0.04} \\
\bottomrule
\end{tabular}
\end{adjustbox}
\end{table}

%% file: tables/rq1_detection.tex
\begin{table}[!htbp]
\centering
\small
\caption{Deepfake detection EER (\%) on the \textit{\name{}-Bench}: Base after forensic question tuning (QT) and after grounded rationale tuning (final), and Plus. Up to 10,000 trials per set, the same trials for every column; full-set results on the public protocols are in Table~\ref{tab:arena}.}
\label{tab:rq1-det}
\begin{adjustbox}{max width=\linewidth}
\begin{tabular}{lccc}
\toprule
Test set & Base (QT) & Base & Plus \\
\midrule
ASVspoof 5 & 2.39 & \textbf{2.23} & 3.41 \\
In-the-Wild & 1.51 & \textbf{1.25} & 1.83 \\
ASVspoof 2021 LA & 3.40 & 5.17 & \textbf{3.18} \\
ASVspoof 2021 DF & 1.14 & 1.07 & \textbf{0.76} \\
ASVspoof 2019 LA & 0.47 & 0.56 & \textbf{0.18} \\
SpoofCeleb & 0.30 & 0.20 & \textbf{0.10} \\
CodecFake+ CoSG & 4.51 & 4.84 & \textbf{2.34} \\
CodecFake+ & 2.24 & 3.95 & \textbf{0.65} \\
DFADD & \textbf{0.00} & \textbf{0.00} & \textbf{0.00} \\
LibriSeVoc & \textbf{0.00} & \textbf{0.00} & \textbf{0.00} \\
\midrule
LA19 aug. & 5.80 & 5.93 & \textbf{5.65} \\
LA19 cond. & 5.00 & 5.00 & \textbf{3.87} \\
DFADD aug. & 2.64 & 2.64 & \textbf{2.13} \\
DFADD cond. & \textbf{1.44} & 1.84 & 1.60 \\
LibriSeVoc aug. & 10.28 & 10.31 & \textbf{7.72} \\
LibriSeVoc cond. & 6.33 & 5.67 & \textbf{4.33} \\
\bottomrule
\end{tabular}
\end{adjustbox}
\\[2pt]\parbox{\linewidth}{\footnotesize aug.: augmented copy (noise, reverberation, codecs); cond.: recorded under a logged condition.}
\end{table}

%% file: tables/rq1_source.tex
\begin{table}[t]
\centering
\small
\caption{Source tasks. Source verification: EER (\%) for same vs.\ different generator on spoof--spoof pairs (top: clean; middle: augmented / logged-condition copies). Bottom: attribution and recording-condition accuracy (\%). Frozen probe: logistic-regression pair probe on the frozen anti-spoofing encoder DF-Arena-1B (learned layer mix), trained on the Base training lists and tested on the same pairs (Appendix~\ref{app:probe}); --: not applicable. Zero-shot audio LLMs: Appendix~\ref{tab:app-zeroshot}.}
\label{tab:rq1-src}
\begin{adjustbox}{max width=\linewidth}
\begin{tabular}{lccc}
\toprule
Test set & Frozen probe & Base & Plus \\
\midrule
MLAAD seen & 29.0 & 17.23 & 7.68 \\
MLAAD unseen model & 26.5 & 13.80 & 6.17 \\
MLAAD unseen arch. & 24.8 & 20.25 & 15.00 \\
SpoofCeleb & 21.9 & 11.25 & 12.20 \\
STOPA & 17.9 & 20.88 & 13.55 \\
ASVspoof 2019 LA & 13.6 & 8.22 & 7.62 \\
ASVspoof 5 & 22.2 & 23.27 & 20.65 \\
\midrule
MLAAD aug. & 31.1 & 30.33 & 23.43 \\
MLAAD cond. & 37.9 & 36.70 & 29.95 \\
STOPA aug. & 33.2 & 29.00 & 22.97 \\
STOPA cond. & 36.5 & 31.70 & 25.55 \\
\midrule
Generator ID acc. (MLAAD seen) & -- & 95.4 & 96.8 \\
Acoustic model acc. (unseen vocoder) & -- & 84.7 & 69.8 \\
Recording condition acc. & -- & 75.9 & 81.3 \\
\bottomrule
\end{tabular}
\end{adjustbox}
\end{table}

%% file: tables/app_zeroshot.tex
\begin{table}[!htbp]
\centering
\small
\caption{Zero-shot general audio LLMs on the \textit{\name{}-Bench}: balanced accuracy (\%) on 1,000 items per set drawn from the test items of Table~\ref{tab:rq1-det} and Table~\ref{tab:app-src}, audio as \name{} hears it. Every question lists its options; greedy decoding, the first option named counts (parse rate $\geq$99.7\%). Base and Plus: the most likely answer sentence on the same items. Pairs are presented as one clip, recording A, 1\,s of silence and recording B. Qwen3-Omni-30B-A3B \citep{qwen3omni}, MiniCPM-o 4.5 \citep{minicpmo45}, Audio Flamingo 3 \citep[AF3;][]{audioflamingo3}, Kimi-Audio-7B \citep{kimiaudio}, Step-Audio-2-mini \citep{stepaudio2}, MiMo-Audio-7B \citep{mimoaudio}.}
\label{tab:app-zeroshot}
\begin{adjustbox}{max width=\linewidth}
\begin{tabular}{lcccccccc}
\toprule
Test set & Base & Plus & Qwen3-Omni & MiniCPM-o & AF3 & Kimi & Step-2 & MiMo \\
\midrule
\multicolumn{9}{@{}l}{\textit{Genuine or spoofed? (chance 50)}} \\
ASVspoof 2021 DF & \textbf{99.7} & \textbf{99.7} & 90.2 & 76.7 & 59.9 & 72.5 & 55.8 & 75.4 \\
ASVspoof 2021 LA & 94.4 & \textbf{96.1} & 78.0 & 70.3 & 59.2 & 62.5 & 54.6 & 72.9 \\
ASVspoof 5 & \textbf{97.5} & 97.1 & 76.0 & 75.8 & 63.9 & 69.2 & 54.8 & 71.6 \\
CodecFake+ CoSG & 90.9 & \textbf{97.6} & 50.2 & 48.6 & 48.9 & 57.9 & 50.2 & 49.8 \\
CodecFake+ & \textbf{90.4} & 89.7 & 52.7 & 59.4 & 50.1 & 65.6 & 50.0 & 52.2 \\
DFADD & \textbf{99.7} & 95.9 & 68.4 & 65.8 & 50.1 & 75.0 & 50.2 & 63.0 \\
DFADD aug. & 96.3 & \textbf{97.2} & 63.5 & 56.7 & 51.0 & 64.8 & 51.2 & 60.7 \\
DFADD cond. & \textbf{98.0} & \textbf{98.0} & 64.0 & 55.7 & 48.8 & 63.6 & 50.7 & 59.9 \\
In-the-Wild & \textbf{98.7} & 97.7 & 64.0 & 61.6 & 49.3 & 66.6 & 52.0 & 58.0 \\
ASVspoof 2019 LA & 99.7 & \textbf{99.8} & 77.8 & 70.1 & 57.6 & 63.9 & 53.6 & 68.5 \\
LA19 aug. & 92.6 & \textbf{93.4} & 70.0 & 54.9 & 55.6 & 64.5 & 53.9 & 63.4 \\
LA19 cond. & 93.5 & \textbf{94.9} & 69.8 & 57.0 & 54.4 & 65.5 & 53.6 & 60.9 \\
LibriSeVoc & \textbf{100.0} & \textbf{100.0} & 52.4 & 57.9 & 50.0 & 55.5 & 49.8 & 50.3 \\
LibriSeVoc aug. & 89.9 & \textbf{92.8} & 52.8 & 56.7 & 49.8 & 54.7 & 49.9 & 49.8 \\
LibriSeVoc cond. & 93.4 & \textbf{94.4} & 54.6 & 58.5 & 50.5 & 57.9 & 49.8 & 51.1 \\
SpoofCeleb & \textbf{99.9} & \textbf{99.9} & 51.2 & 58.3 & 51.0 & 57.2 & 53.3 & 51.5 \\
Mean & 95.9 & \textbf{96.5} & 64.7 & 61.5 & 53.1 & 63.6 & 52.1 & 59.9 \\
\midrule
\multicolumn{9}{@{}l}{\textit{Same source? (chance 50)}} \\
ASVspoof 5 & 74.0 & \textbf{80.1} & 55.0 & 52.3 & 51.4 & 51.5 & 50.4 & 54.2 \\
ASVspoof 2019 LA & 91.0 & \textbf{91.3} & 54.5 & 52.2 & 50.9 & 50.7 & 52.2 & 50.5 \\
MLAAD seen & 85.1 & \textbf{93.6} & 49.4 & 53.0 & 48.1 & 51.1 & 50.1 & 52.5 \\
MLAAD unseen arch. & 81.1 & \textbf{86.4} & 53.1 & 58.4 & 45.2 & 47.3 & 50.2 & 52.3 \\
MLAAD unseen model & 86.6 & \textbf{89.9} & 51.4 & 47.7 & 46.8 & 52.1 & 49.6 & 49.4 \\
MLAAD aug. & 68.9 & \textbf{76.4} & 58.3 & 57.2 & 47.9 & 56.1 & 50.8 & 55.5 \\
MLAAD cond. & 64.1 & \textbf{69.6} & 48.2 & 54.3 & 45.6 & 53.6 & 49.9 & 49.0 \\
SpoofCeleb & 87.1 & \textbf{89.8} & 52.5 & 52.2 & 48.4 & 50.6 & 50.1 & 54.8 \\
STOPA & 80.0 & \textbf{84.4} & 51.6 & 51.0 & 50.2 & 50.7 & 48.1 & 52.9 \\
STOPA aug. & 71.4 & \textbf{75.6} & 50.2 & 54.1 & 48.7 & 50.1 & 51.9 & 52.3 \\
STOPA cond. & 67.7 & \textbf{73.9} & 50.6 & 53.8 & 48.7 & 49.6 & 50.6 & 50.6 \\
Mean & 77.9 & \textbf{82.8} & 52.3 & 53.3 & 48.4 & 51.2 & 50.4 & 52.2 \\
\midrule
\multicolumn{9}{@{}l}{\textit{Recording condition (5--8 options)}} \\
DFADD aug. & 77.0 & \textbf{86.1} & 18.9 & 17.1 & 14.2 & 14.2 & 18.1 & 18.4 \\
DFADD cond. & 81.7 & \textbf{86.0} & 25.0 & 22.8 & 17.3 & 21.8 & 16.6 & 21.9 \\
LA19 aug. & 76.5 & \textbf{80.8} & 20.5 & 15.4 & 12.8 & 14.7 & 19.0 & 18.0 \\
LA19 cond. & 81.5 & \textbf{85.1} & 24.4 & 21.5 & 18.1 & 21.1 & 17.2 & 22.6 \\
MLAAD cond. & 77.5 & \textbf{82.2} & 22.5 & 22.2 & 17.9 & 21.5 & 17.1 & 21.3 \\
STOPA cond. & 67.5 & \textbf{77.7} & 24.0 & 20.9 & 18.1 & 18.7 & 18.6 & 20.3 \\
Mean & 76.9 & \textbf{83.0} & 22.6 & 20.0 & 16.4 & 18.7 & 17.8 & 20.4 \\
\bottomrule
\end{tabular}
\end{adjustbox}
\end{table}

%% file: tables/app_source_1b.tex
\begin{table}[!htbp]
\centering
\small
\caption{Source tasks of Base after forensic question tuning (QT) and after grounded rationale tuning (final), and Plus (same sets as Table~\ref{tab:rq1-src}): EER (\%) of same-source verification on pairs, accuracy (\%) in the last three rows. MLAAD seen / unseen model / unseen arch.: pairs of generators seen in training, unseen models of a seen architecture, unseen architectures.}
\label{tab:app-src}
\begin{adjustbox}{max width=\linewidth}
\begin{tabular}{lccc}
\toprule
Test set & Base (QT) & Base & Plus \\
\midrule
MLAAD seen & 17.38 & 17.23 & \textbf{7.68} \\
MLAAD unseen model & 14.77 & 13.80 & \textbf{6.17} \\
MLAAD unseen arch. & 21.35 & 20.25 & \textbf{15.00} \\
SpoofCeleb & 11.50 & \textbf{11.25} & 12.20 \\
STOPA & 20.05 & 20.88 & \textbf{13.55} \\
ASVspoof 2019 LA & 8.68 & 8.22 & \textbf{7.62} \\
ASVspoof 5 & 23.52 & 23.27 & \textbf{20.65} \\
\midrule
MLAAD aug. & 29.37 & 30.33 & \textbf{23.43} \\
MLAAD cond. & 35.40 & 36.70 & \textbf{29.95} \\
STOPA aug. & 28.47 & 29.00 & \textbf{22.97} \\
STOPA cond. & 32.20 & 31.70 & \textbf{25.55} \\
\midrule
Generator ID acc. (MLAAD seen) & 96.3 & 95.4 & \textbf{96.8} \\
Acoustic model acc. (unseen vocoder) & 82.3 & \textbf{84.7} & 69.8 \\
Recording condition acc. & 77.9 & 75.9 & \textbf{81.3} \\
\bottomrule
\end{tabular}
\end{adjustbox}
\end{table}

%% file: tables/analysis_composition.tex
\begin{table}[!htbp]
\centering
\small
\caption{Compositional generalization (SpoofCeleb systems A17 and A18: an acoustic model seen in training with a vocoder never seen with it): linear-probe accuracy of the acoustic model on encoder states and pooled audio tokens, and the model's answer (LLM rows after question tuning, as in Table~\ref{tab:rq2}(b); Base: final; chance 0.50).}
\label{tab:comp}
\begin{adjustbox}{max width=\linewidth}
\begin{tabular}{lccc}
\toprule
Model & Encoder probe & Token probe & LLM answer \\
\midrule
Qwen2.5-1.5B & 0.87 & 0.66 & 0.52 \\
MiniCPM5-2B & 0.84 & 0.64 & 0.86 \\
Qwen3-1.7B & 0.90 & 0.75 & 0.64 \\
\rowcolor{oursbg}Base & -- & -- & 0.85 \\
\bottomrule
\end{tabular}
\end{adjustbox}
\end{table}

%% file: tables/rq2_main.tex
\begin{table}[!htbp]
\centering
\small
\caption{Components. (a) Representation ladder after perception alignment (LLM frozen; Qwen2.5-1.5B up to R6): each row changes one factor of the row above. R0 trains on Qwen3-ASR features of layers 8 and 11 computed once on 4 s windows; R1 runs the encoder during training, so augmentation is drawn anew for every example; R4 adds the one-class loss and the full ASVspoof~5 training split; R6 adds narrowband codecs to augmentation; R8 applies post-processing to both classes. SV clean / deg.: mean EER (\%) of same-source verification over the 7 clean and 4 degraded sets of Table~\ref{tab:app-src}; Det. clean: mean detection EER over LA19, ASVspoof~5, SpoofCeleb and CoSG; Det. deg.: over the 6 degraded sets of Table~\ref{tab:rq1-det}; ASV5: ASVspoof~5. (b) Only the LLM changes (question tuning, same recipe; Qwen2.5-1.5B \citep{qwen25}, MiniCPM5-2B \citep{minicpm5}, Qwen3-1.7B \citep{qwen3}, Qwen3.5-4B \citep{qwen35}): mean EER (\%) over the same SV sets and over the first 8 detection sets of Table~\ref{tab:rq1-det}; localization IoU (HAD, PartialEdit) and acoustic-model (AM) accuracy under unseen vocoders after a rationale-tuning run of the same recipe, which was not run for Qwen3.5-4B (--). Hybrid: linear-attention and full-attention layers. Shaded: the configuration used.}
\label{tab:rq2}
\begin{adjustbox}{max width=\linewidth}
\begin{tabular}{lccccc}
\toprule
\textit{(a) Variant} & SV clean & SV deg. & Det. clean & Det. deg. & ASV5 \\
\midrule
R0 precomputed layers 8, 11; 4 s & 19.02 & 32.17 & 4.97 & 9.00 & 9.56 \\
R1 + encoder in the loop & 17.04 & 33.70 & 4.28 & 9.80 & 8.06 \\
R2 + 8 s window & 16.90 & 33.80 & 4.45 & 8.35 & 8.54 \\
R3 + 25-layer fusion & 16.59 & 31.95 & 3.77 & 8.57 & 5.80 \\
R4 + one-class loss, full ASV5 & 17.99 & 33.51 & 3.78 & 8.39 & 5.80 \\
R5 + encoder DoRA & 16.41 & 32.16 & 2.89 & 6.33 & 5.95 \\
R6 + narrowband codecs & \textbf{16.38} & 31.98 & 2.96 & \textbf{5.91} & 4.27 \\
R7 LLM $\to$ MiniCPM5-2B & 17.00 & \textbf{29.99} & 2.61 & 6.20 & 3.43 \\
\rowcolor{oursbg}R8 + post-processing, both classes & 18.19 & 33.87 & \textbf{1.84} & 6.62 & \textbf{3.10} \\
\midrule
\textit{(b) LLM, question tuning} & SV clean & SV deg. & Det. & Loc.\ IoU & AM acc. \\
\midrule
Qwen2.5-1.5B & \textbf{14.61} & 30.59 & 2.05 & 0.36 & 0.77 \\
\rowcolor{oursbg}MiniCPM5-2B & 16.33 & \textbf{28.98} & 2.31 & \textbf{0.61} & \textbf{0.84} \\
Qwen3-1.7B & 15.64 & 31.36 & 2.22 & 0.31 & 0.76 \\
Qwen3.5-4B (hybrid) & 20.15 & 37.51 & \textbf{1.85} & -- & -- \\
\bottomrule
\end{tabular}
\end{adjustbox}
\end{table}

%% file: tables/rq2_shortcut.tex
\begin{table}[!htbp]
\centering
\small
\caption{Shortcut probes (perception alignment, 4,000 utterances per set): error rates (\%) after a 2 kHz high-pass (spoof judged bona fide; HP), 0.5 s digital-silence padding (Pad) and +10 dB gain with clipping (Gain; bona fide judged spoof), on In-the-Wild (ITW), ASVspoof~5 (ASV5) and ASVspoof 2019 LA (LA19). Labelled: the answer also names the post-processing.}
\label{tab:rq2-short}
\begin{adjustbox}{max width=\linewidth}
\begin{tabular}{lccccc}
\toprule
Training & HP ITW & HP ASV5 & HP LA19 & Pad ASV5 & Gain ASV5 \\
\midrule
No post-processing & 11.9 & 11.6 & 26.8 & 12.4 & 15.6 \\
\rowcolor{oursbg}+ post-processing & 3.7 & \textbf{1.7} & \textbf{7.4} & \textbf{4.0} & \textbf{8.4} \\
+ post-processing, labelled & \textbf{2.2} & 1.9 & 8.9 & 4.4 & \textbf{8.4} \\
\bottomrule
\end{tabular}
\end{adjustbox}
\end{table}

%% file: tables/rq3.tex
\begin{table}[!htbp]
\centering
\small
\caption{Source supervision: multi-task model (A) vs.\ authenticity questions only, with as many authenticity questions as A (B) or as many training steps (B$'$, about 5$\times$ A's authenticity questions). $D = \mathrm{EER}_{B} - \mathrm{EER}_A$ ($>0$: A is better), paired bootstrap (2,000 resamples, stratified by class), 95\% interval; decision rule fixed before the results (equivalence margin $\pm 0.5$).}
\label{tab:rq3}
\begin{adjustbox}{max width=\linewidth}
\begin{tabular}{lccclccl}
\toprule
Test set & EER A & EER B & $D$ [95\% CI] & Verdict & EER B$'$ & $D'$ [95\% CI] & Verdict \\
\midrule
ASVspoof 5 (all 30,005) & 2.44 & 3.57 & 1.13 [0.67, 1.81] & helps & 2.04 & -0.40 [-0.80, 0.00] & inconclusive \\
ASVspoof 2021 DF & 1.14 & 1.12 & -0.02 [-0.45, 0.66] & inconclusive & 0.85 & -0.29 [-0.58, 0.70] & inconclusive \\
ASVspoof 2021 LA & 3.40 & 1.77 & -1.63 [-2.24, -1.19] & hurts & 2.09 & -1.31 [-1.88, -1.01] & hurts \\
CodecFake+ CoSG & 4.51 & 7.29 & 2.78 [1.61, 3.95] & helps & 5.29 & 0.78 [-0.11, 1.61] & inconclusive \\
CodecFake+ & 2.24 & 8.40 & 6.16 [5.08, 7.34] & helps & 12.62 & 10.38 [8.73, 11.59] & helps \\
DFADD & 0.00 & 0.00 & 0.00 [0.00, 0.00] & equivalent & 0.00 & 0.00 [0.00, 0.00] & equivalent \\
DFADD aug. & 2.64 & 3.44 & 0.80 [0.00, 1.46] & inconclusive & 2.53 & -0.11 [-0.67, 0.53] & inconclusive \\
DFADD cond. & 1.44 & 2.13 & 0.69 [0.15, 1.18] & helps & 1.71 & 0.27 [-0.24, 0.69] & inconclusive \\
In-the-Wild & 1.51 & 1.70 & 0.19 [-0.05, 0.43] & equivalent & 1.23 & -0.28 [-0.48, -0.01] & hurts \\
ASVspoof 2019 LA & 0.47 & 0.19 & -0.28 [-0.46, 0.00] & equivalent & 0.57 & 0.10 [-0.05, 0.37] & equivalent \\
LA19 aug. & 5.80 & 6.60 & 0.80 [0.02, 1.53] & helps & 6.53 & 0.73 [-0.05, 1.33] & inconclusive \\
LA19 cond. & 5.00 & 6.00 & 1.00 [0.20, 1.60] & helps & 5.67 & 0.67 [-0.07, 1.25] & inconclusive \\
LibriSeVoc & 0.00 & 0.33 & 0.33 [0.00, 1.06] & inconclusive & 0.00 & 0.00 [0.00, 0.00] & equivalent \\
LibriSeVoc aug. & 10.28 & 16.67 & 6.39 [4.25, 8.61] & helps & 8.33 & -1.94 [-3.39, -0.44] & hurts \\
LibriSeVoc cond. & 6.33 & 7.69 & 1.36 [0.33, 3.00] & helps & 5.08 & -1.25 [-2.39, 0.00] & inconclusive \\
SpoofCeleb & 0.30 & 0.52 & 0.22 [0.00, 0.59] & inconclusive & 0.21 & -0.09 [-0.28, 0.18] & equivalent \\
\bottomrule
\end{tabular}
\end{adjustbox}
\end{table}

%% file: tables/app_grounded.tex
\begin{table}[!htbp]
\centering
\small
\caption{Grounded detection reasoning of \name{}-Base, all measures (500 utterances per set). Acc.: decision after the model's own analysis (\%). Agree: agreement with the one-step answer (\%). BW (bandwidth class), SNR, Level (speech level), Clip (clipping): share of correct observation statements. IoU of stated spans and F1 of the per-token head on partially spoofed sets. Flip / Para.: share of decisions that follow an opposite localization, in the edit form used in training / a paraphrased form. +proc: copies with post-processing.}
\label{tab:app-grounded}
\begin{adjustbox}{max width=\linewidth}
\begin{tabular}{lcccccccccc}
\toprule
Set & Acc. & Agree & BW & SNR & Level & Clip & IoU & Tok.\ F1 & Flip & Para. \\
\midrule
LA19 & 99.6 & 99.6 & 0.71 & 0.97 & 0.99 & 0.86 & -- & 0.95 & 1.00 & 1.00 \\
ASV5 & 97.0 & 98.2 & 0.77 & 0.88 & 0.99 & 0.73 & -- & 0.98 & 1.00 & 1.00 \\
ITW & 98.8 & 98.6 & 0.64 & 0.94 & 0.96 & 0.99 & -- & 0.99 & 1.00 & 1.00 \\
CoSG & 97.6 & 93.8 & 0.77 & 0.92 & 1.00 & 0.98 & -- & 0.97 & 1.00 & 1.00 \\
MLAAD & 97.4 & 87.2 & 0.72 & 0.95 & 0.99 & 0.99 & -- & -- & 1.00 & 0.96 \\
PartialSpoof & 91.2 & 96.6 & 0.79 & 0.98 & 0.99 & 1.00 & 0.25 & 0.74 & 1.00 & 1.00 \\
HAD & 99.0 & 99.6 & 0.91 & 0.99 & 1.00 & 0.87 & 0.61 & 0.91 & 1.00 & 1.00 \\
PartialEdit & 99.6 & 78.8 & 0.68 & 0.98 & 0.99 & 1.00 & 0.58 & 0.79 & 1.00 & 1.00 \\
LlamaPS & 92.2 & 91.2 & 0.82 & 0.94 & 1.00 & 0.94 & 0.31 & 0.82 & 1.00 & 0.99 \\
LA19+proc & 99.0 & 98.6 & 0.75 & 0.98 & 0.99 & 0.87 & -- & 0.95 & 1.00 & 1.00 \\
ITW+proc & 98.0 & 97.6 & 0.67 & 0.93 & 0.95 & 0.95 & -- & 0.98 & 1.00 & 1.00 \\
\bottomrule
\end{tabular}
\end{adjustbox}
\end{table}

%% file: tables/rq4_loc.tex
\begin{table}[!htbp]
\centering
\small
\caption{Localization: IoU of the spans stated in the reasoning vs.\ F1 of the per-token head (synthesized class, 0.16 s grid) on the same utterances of HAD, PartialEdit (PEdit), PartialSpoof (PSpoof) and LlamaPartialSpoof (LlamaPS). LLM rows: the attribute-first rationale-tuning runs of Table~\ref{tab:rq4-svr}.}
\label{tab:rq4-loc}
\begin{adjustbox}{max width=\linewidth}
\begin{tabular}{lcccccccc}
\toprule
Model & IoU HAD & PEdit & PSpoof & LlamaPS & F1 HAD & PEdit & PSpoof & LlamaPS \\
\midrule
Qwen2.5-1.5B & 0.35 & 0.36 & 0.24 & 0.30 & 0.89 & 0.74 & 0.73 & 0.81 \\
Qwen3-1.7B & 0.25 & 0.38 & 0.19 & 0.30 & 0.90 & 0.76 & 0.74 & 0.80 \\
MiniCPM5-2B & \textbf{0.64} & \textbf{0.59} & \textbf{0.27} & \textbf{0.31} & \textbf{0.91} & 0.77 & \textbf{0.76} & 0.81 \\
\rowcolor{oursbg}Base & 0.61 & 0.58 & 0.25 & \textbf{0.31} & \textbf{0.91} & \textbf{0.79} & 0.74 & \textbf{0.82} \\
\bottomrule
\end{tabular}
\end{adjustbox}
\end{table}

%% file: tables/app_agree.tex
\begin{table}[!htbp]
\centering
\small
\caption{Reasoning decision vs.\ direct answer on the same 500 utterances per set (\%): agreement, accuracy of the reasoning decision and of the direct answer.}
\label{tab:app-agree}
\begin{adjustbox}{max width=\linewidth}
\begin{tabular}{lcccccc}
\toprule
Set & Base agree & Base reason. & Base direct & Plus agree & Plus reason. & Plus direct \\
\midrule
LA19 & 99.6 & 99.6 & 99.2 & 99.0 & 99.0 & 100.0 \\
ASV5 & 98.2 & 97.0 & 97.6 & 98.6 & 98.8 & 97.4 \\
ITW & 98.6 & 98.8 & 99.4 & 99.6 & 98.0 & 98.4 \\
CoSG & 93.8 & 97.6 & 91.8 & 98.4 & 98.2 & 97.8 \\
MLAAD & 87.2 & 97.4 & 85.8 & 99.6 & 99.4 & 99.4 \\
PartialSpoof & 96.6 & 91.2 & 90.2 & 90.0 & 89.2 & 80.0 \\
HAD & 99.6 & 99.0 & 99.0 & 50.4 & 97.6 & 48.4 \\
PartialEdit & 78.8 & 99.6 & 78.4 & 94.8 & 99.6 & 94.4 \\
LlamaPS & 91.2 & 92.2 & 87.4 & 96.2 & 90.2 & 88.4 \\
LA19+proc & 98.6 & 99.0 & 98.4 & -- & -- & -- \\
ITW+proc & 97.6 & 98.0 & 98.8 & -- & -- & -- \\
\bottomrule
\end{tabular}
\end{adjustbox}
\end{table}

%% file: tables/rq4_svr.tex
\begin{table}[!htbp]
\centering
\small
\caption{Source-verification reasoning: verdict EER (\%) after the model's own analysis; REV acc.: accuracy on REVERSAL pairs (different systems sharing every attribute class); Direct: EER of the one-step answer on the same pairs (LA19 cond.). Attribute-first / fingerprint-first: two earlier rationale-tuning runs (MiniCPM5-2B unless stated, without counterfactual examples) that differ only in whether the attribute comparison or the fingerprint verdict comes first; Base and Plus state the fingerprint verdict first.}
\label{tab:rq4-svr}
\begin{adjustbox}{max width=\linewidth}
\begin{tabular}{lcccccc}
\toprule
Format & SpoofCeleb & STOPA & CoSG & LA19 cond. & REV acc. & Direct \\
\midrule
Attribute-first & 23.40 & 11.75 & 24.20 & 31.70 & 0.31 & 19.50 \\
Fingerprint-first & 22.80 & 10.80 & 26.60 & 24.95 & 0.67 & \textbf{18.35} \\
\rowcolor{oursbg}Base & 20.20 & 13.65 & \textbf{21.60} & 29.65 & 0.42 & 28.45 \\
\rowcolor{oursbg}Plus & \textbf{18.40} & 14.35 & 21.80 & 24.55 & 0.75 & 20.20 \\
Attribute-first, Qwen2.5-1.5B & 19.40 & \textbf{10.50} & 27.00 & \textbf{22.75} & \textbf{0.76} & 20.50 \\
\bottomrule
\end{tabular}
\end{adjustbox}
\end{table}

%% file: sections/conclusion.tex
\section{Conclusion}
\label{sec:conclusion}
Speech forensics can be cast as scored question answering with grounded and faithful rationales. \textit{\name{}} reads every
answer as a score, writes rationales whose statements can be checked against the signal, and is trained so that its
decision follows the evidence it states. With one set of weights it is the best open-weight detector on Speech-DF-Arena
and SpoofRadar, verifies the source of recordings from unseen generators, attributes generators and describes
recording channels. Controlled studies show where to invest: the representation and how it is trained decide what is
detected, the LLM decides how well recordings are compared and how evidence is worded, and supervision that
distinguishes how audio was produced adds what more binary labels cannot, at least for neural-codec fakes. Its
rationales catch partial fakes that one score misses and determine the decision even under edits never seen in
training. We release both models, the inference code and a demo, and the evaluation protocol for the questions beyond
detection.

%% file: sections/appendix.tex
\section{Test sets and licenses}
\label{app:data}
\label{app:tasks}

\begin{figure}[!htbp]
\centering
\figcard{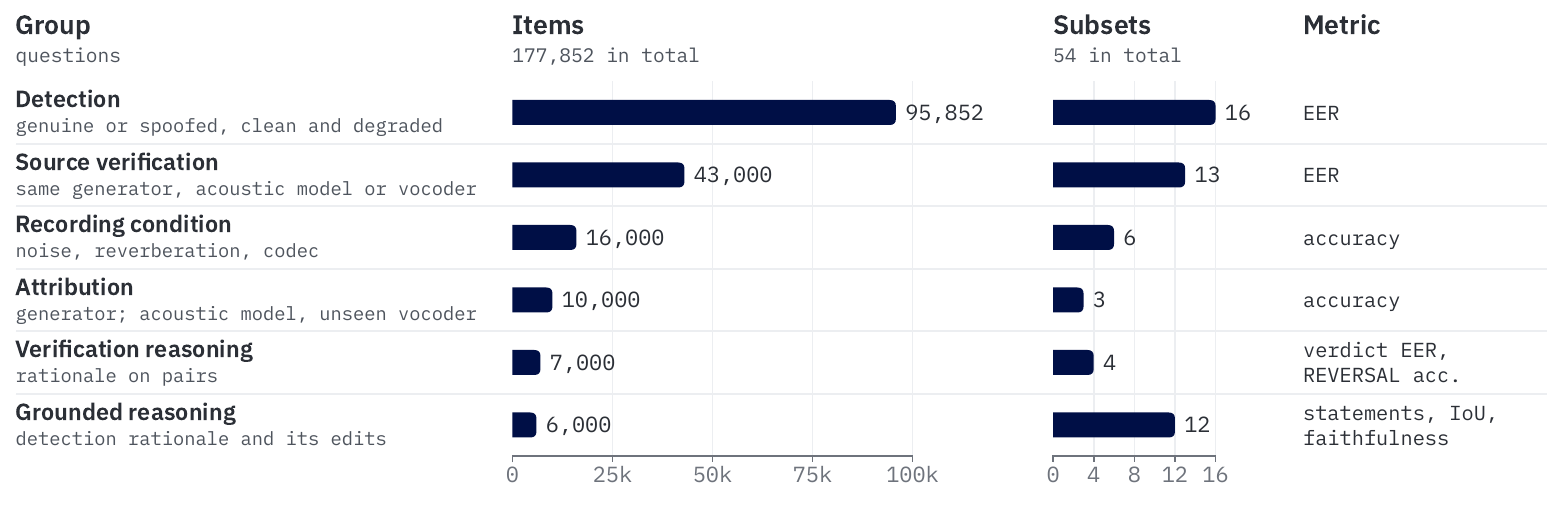}
\caption{\textbf{VeriSpoof-Bench} by group: items and subsets per group, and the metric of each group. Released with
item lists, labels, prompts, scoring scripts and the per-item outputs of both models, without audio.}
\label{fig:bench}
\end{figure}

Figure~\ref{fig:bench} summarizes \textit{VeriSpoof-Bench}; its dataset card lists all 54 subsets with their sources, class
balance and the generators or conditions unseen in training.

Table~\ref{tab:app-card} gives the size of every detection test set of the SpoofRadar protocol, whether the training
split of the same corpus is used by Base and Plus, and the EER of both models. Table~\ref{tab:app-license} lists the
licenses of the corpora in the \textit{\name{}-Bench}. For ASVspoof 2021 DF and In-the-Wild the distributors' statements
conflict, and the table gives both. ADD 2022 / 2023 and Codecfake (CC BY-NC-ND 4.0) and Fake-or-Real and SONAR (no
license found) are not part of the released lists; they enter only the training data of Plus (ADD 2022, Codecfake,
Fake-or-Real) and the comparison on the public protocols. On the 14 common sets alone, Plus has 2.19\% pooled EER under
the SpoofRadar lists (2.14\% in Table~\ref{tab:arena}, which uses the full LibriSeVoc list and 4,528 Fake-or-Real
trials).

\input{tables/app_card}
\input{tables/app_license}

\section{Training settings}
\label{app:hparams}
Table~\ref{tab:app-hparams} gives the settings of the three stages for both models. Plus runs perception alignment for six balanced epochs over its 211 sources
and keeps the step counts of the other two stages.

\input{tables/app_hparams}

\section{Encoder probes}
\label{app:probe}
\label{app:shortcut}
\label{app:rq2}
Figure~\ref{fig:rq2}a compares seven encoders under an earlier protocol: mean and standard deviation over frames of
each layer on a fixed 4.04\,s window, detection by logistic regression on up to 8,000 bona fide and 8,000 spoofed
utterances per training list, source verification by an MLP on $[|a-b|, a \odot b]$ of 120k training pairs, and
earlier test pairs. Table~\ref{tab:app-probe} repeats the comparison of Qwen3-ASR with DF-Arena-1B under the current
protocol: the Base training lists, the test items of the model evaluation and logistic-regression probes. The
frozen-probe column of Table~\ref{tab:rq1-src} is the DF-Arena-1B pair probe with a learned layer mix.

\input{tables/app_probe}

\section{Questions and example outputs}
\label{app:examples}
\label{app:loc}
\label{app:reason}
\label{app:rq3}
\label{app:zeroshot}
\label{app:arena}
\label{app:spoofradar}
\begin{figure}[!htbp]
\centering
\figcard{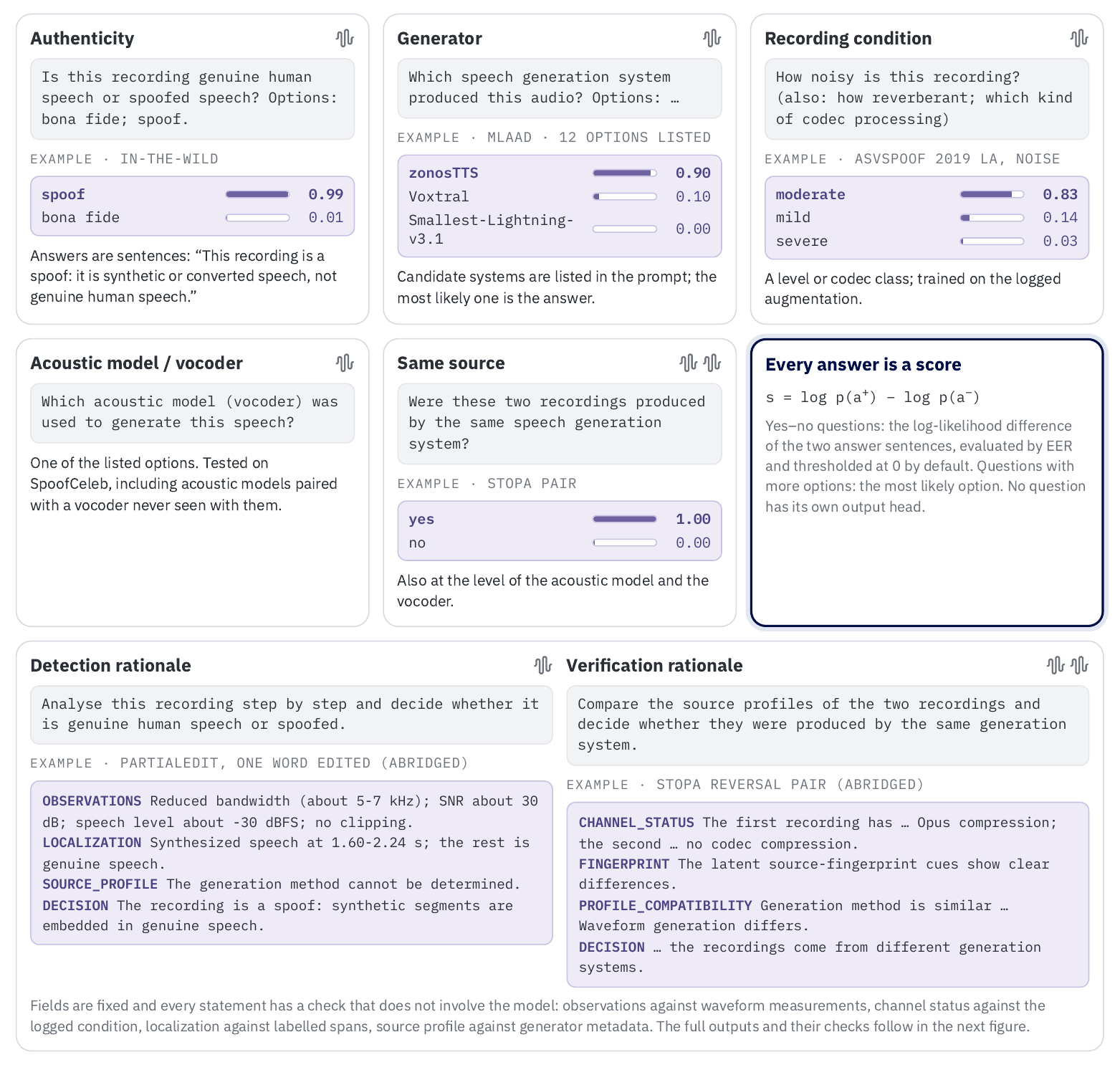}
\caption{\textbf{The questions.} Each card gives the question template, an example answer of \textit{\name{}-Base} with the
probabilities of its options, and the answer form. One-step answers are read as scores (orange card); rationales are
written in fixed fields, abridged here.}
\label{fig:questions}
\end{figure}

Figure~\ref{fig:questions} shows every question \textit{\name{}} answers, with its template (one of several paraphrases) and an
example answer of \textit{\name{}-Base}; the probabilities are those of Figure~\ref{fig:overview}. Figure~\ref{fig:app-examples} shows two full outputs with their checks: a PartialEdit fake missed by the one-step answer
and caught by the rationale, which also contains a bandwidth statement that fails its check; and a REVERSAL pair
decided correctly by the fingerprint verdict although the stated attributes are wrong, an error the checks expose.

\input{tables/app_examples}

%% file: tables/app_card.tex
\begin{table}[!htbp]
\centering
\small
\caption{Detection test sets of the SpoofRadar v1.0 protocol (the 14 common sets, with LibriSeVoc restricted to its test part, plus XMAD-Bench and SpoofCeleb): trials, and whether the training split of the same corpus is used by Base and Plus, and EER (\%) of both models; Total row: pooled EER. LA19: ASVspoof 2021 has no training split, and the ASVspoof 2019 LA one is used; ADD22: ADD 2022 train / dev. $^\ast$ AISHELL-3 bona fide speech, 1,001 of the XMAD trials.}
\label{tab:app-card}
\begin{adjustbox}{max width=\linewidth}
\begin{tabular}{lrrrcccc}
\toprule
Set & Trials & Bona fide & Spoof & Base train & Plus train & EER Base & EER Plus \\
\midrule
ASVspoof2019-LA & 71,237 & 7,355 & 63,882 & yes & yes & 0.49 & \textbf{0.22} \\
ASVspoof2021-LA & 148,176 & 14,816 & 133,360 & LA19 & LA19 & 5.48 & \textbf{3.44} \\
ASVspoof2021-DF & 533,928 & 14,869 & 519,059 & LA19 & LA19 & \textbf{0.97} & 1.24 \\
ASVspoof5-Eval & 680,774 & 138,688 & 542,086 & yes & yes & \textbf{2.59} & 3.40 \\
ADD2022-T1 & 109,199 & 31,334 & 77,865 & no & ADD22 & 14.05 & \textbf{8.65} \\
ADD2022-T3 & 126,861 & 20,776 & 106,085 & no & ADD22 & 3.87 & \textbf{1.18} \\
ADD2023-R1 & 111,976 & 80,000 & 31,976 & no & ADD22 & 5.01 & \textbf{0.50} \\
ADD2023-R2 & 118,477 & 87,500 & 30,977 & no & ADD22 & 10.61 & \textbf{2.41} \\
Codecfake & 304,241 & 92,596 & 211,645 & no & yes & 4.97 & \textbf{0.12} \\
LibriSeVoc & 18,487 & 2,641 & 15,846 & yes & yes & \textbf{0.00} & \textbf{0.00} \\
DFADD & 3,755 & 755 & 3,000 & yes & yes & \textbf{0.00} & \textbf{0.00} \\
SONAR & 3,948 & 2,274 & 1,674 & no & no & 5.19 & \textbf{1.92} \\
In-the-Wild & 31,779 & 19,963 & 11,816 & no & no & \textbf{1.20} & 1.90 \\
FoR & 4,634 & 2,264 & 2,370 & no & yes & 5.31 & \textbf{0.09} \\
XMAD-Bench & 102,220 & 51,110 & 51,110 & no & AISHELL-3$^\ast$ & \textbf{0.24} & 1.33 \\
SpoofCeleb & 91,130 & 9,113 & 82,017 & yes & yes & 0.20 & \textbf{0.04} \\
\midrule
Total & 2,460,822 & 576,054 & 1,884,768 &  &  & 4.23 & 4.04 \\
\bottomrule
\end{tabular}
\end{adjustbox}
\end{table}

%% file: tables/app_license.tex
\begin{table}[!htbp]
\centering
\small
\caption{Licenses of the internal-suite corpora as stated by each distributor (checked 31 Oct 2026). We release lists,
labels and scripts, not audio; each subset's lists carry its corpus' terms.}
\label{tab:app-license}
\begin{tabular}{@{}l p{0.43\linewidth} p{0.27\linewidth}@{}}
\toprule
Corpus & License & Terms for our lists \\
\midrule
ASVspoof 2019 LA & ODC-By 1.0 & attribution \\
ASVspoof 2021 LA & ODC-By & attribution \\
ASVspoof 2021 DF & ODbL (Zenodo record) or ODC-By (asvspoof.org) & attribution; share-alike if ODbL \\
ASVspoof~5 & ODC-By 1.0 (bona fide speech CC BY 4.0) & attribution \\
In-the-Wild & Apache-2.0 (distributor) or CC BY-SA 4.0 (authors' card) & attribution; share-alike if CC BY-SA \\
SpoofCeleb & CC BY 4.0, gated & attribution \\
MLAAD & CC BY-NC 4.0, gated & non-commercial, attribution \\
STOPA & CC BY 4.0 & attribution \\
DFADD & MIT & none \\
LibriSeVoc & CC BY-SA 4.0 & share-alike \\
CodecFake+ & MIT & none \\
PartialSpoof, HAD, PartialEdit & CC BY 4.0 & attribution \\
LlamaPartialSpoof & CC BY 4.0 & attribution \\
\bottomrule
\end{tabular}
\end{table}

%% file: tables/app_hparams.tex
\begin{table}[!htbp]
\centering
\small
\caption{Training settings. PA: perception alignment (tokenizer and encoder DoRA, LLM frozen); QT: forensic question
tuning (LoRA and tokenizer, all perception questions); RT: grounded rationale tuning (rationales, counterfactual
examples, replay). Task weights are sampling weights. Family: TTS or voice conversion; profile: the attribute classes
of one recording's source (generation method, acoustic modelling, waveform generation); authenticity pair: the
authenticity of both recordings of a pair; attribute compatibility: whether two recordings from different systems share
one attribute class; condition: recording condition. One-class loss: OC-softmax \citep{ocsoftmax} on the pooled audio tokens, pulling
bona fide speech towards a learned centre; per-token head: a genuine / synthesized classifier on the LLM's hidden
states of the frame tokens.}
\label{tab:app-hparams}
\begin{tabular}{@{}lll@{}}
\toprule
 & \name{}-Base & \name{}-Plus \\
\midrule
Encoder & \multicolumn{2}{l}{Qwen3-ASR \citep{qwen3asr}, 8\,s windows, 25 layers mixed by learned weights,} \\
 & \multicolumn{2}{l}{DoRA \citep{dora} $r{=}16$ (PA only)} \\
Tokenizer & \multicolumn{2}{l}{16 utterance tokens, 52 frame tokens of 0.16\,s, time markers every 0.8\,s} \\
LLM & \multicolumn{2}{l}{MiniCPM5-2B \citep{minicpm5,minicpm4},} \\
 & \multicolumn{2}{l}{LoRA \citep{lora} $r{=}16$, $\alpha{=}32$, dropout 0.05 on q/k/v/o} \\
Losses & \multicolumn{2}{l}{cross-entropy; one-class authenticity loss (0.1); per-token head (0.3, RT)} \\
\midrule
Steps PA / QT / RT & 8k / 20k / 10k & 79.2k (6 balanced epochs) / 20k / 10k \\
Global batch & 32 & 64 (two GPUs, data parallel) \\
Learning rate PA & tokenizer and DoRA $10^{-4}$ & same \\
Learning rate QT & tokenizer $5{\times}10^{-5}$, LoRA $10^{-4}$ & same \\
Learning rate RT & tokenizer $10^{-5}$, LoRA $2{\times}10^{-5}$ & same \\
Warm-up & 5\% & 2\% (PA), 5\% (QT, RT) \\
\midrule
Task weights PA & \multicolumn{2}{l}{authenticity 3 (Plus 7), family 1, generator 2, same generator 3, recording condition 1} \\
Task weights QT & \multicolumn{2}{l}{authenticity 3, family 1, generator 4, acoustic model 1, vocoder 1, profile 1,} \\
 & \multicolumn{2}{l}{same generator 3, same acoustic model 1, same vocoder 1, condition 3, authenticity pair 2} \\
Task weights RT & \multicolumn{2}{l}{detection rationale 5, verification rationale 4, attribute compatibility 1.5, QT replay} \\
Counterfactual & \multicolumn{2}{l}{20\% of rationale examples, localization reversed, only the decision trained} \\
\bottomrule
\end{tabular}
\end{table}

%% file: tables/app_probe.tex
\begin{table}[!htbp]
\centering
\small
\caption{Frozen-encoder probes under the current protocol: the Base training lists (120,000 training pairs, 70,744 training utterances), the test items of the model evaluation (8\,s windows as the model hears them) and logistic-regression probes on per-layer pooled statistics, with a learned layer mix (mix) or the best single layer (Qwen3-ASR: layer 12 for both tasks; DF-Arena-1B: 8 for pairs, 14 for detection). EER (\%). (strata): pairs balanced over SAME, REVERSAL and DIFF.}
\label{tab:app-probe}
\begin{adjustbox}{max width=\linewidth}
\begin{tabular}{lcccc}
\toprule
Test set & Qwen3-ASR mix & best & DF-Arena-1B mix & best \\
\midrule
\multicolumn{5}{@{}l}{\textit{Source verification (pairs)}} \\
ASVspoof 5 & 25.8 & 23.7 & 22.2 & \textbf{21.5} \\
ASVspoof 2019 LA & \textbf{10.4} & 10.7 & 13.6 & 15.3 \\
MLAAD seen & \textbf{28.4} & 30.4 & 29.0 & 33.2 \\
MLAAD unseen arch. & \textbf{19.5} & 22.6 & 24.8 & 23.8 \\
MLAAD unseen model & \textbf{22.2} & 24.6 & 26.5 & 31.8 \\
MLAAD aug. & \textbf{24.2} & 31.3 & 31.1 & 32.5 \\
MLAAD cond. & \textbf{35.8} & 37.1 & 37.9 & 40.1 \\
SpoofCeleb & \textbf{18.9} & 23.5 & 21.9 & 22.7 \\
STOPA & 22.8 & 22.3 & \textbf{17.9} & 23.8 \\
STOPA aug. & 33.6 & \textbf{32.4} & 33.2 & 34.8 \\
STOPA cond. & 37.8 & 37.4 & \textbf{36.5} & 39.8 \\
CoSG (strata) & \textbf{25.8} & 26.4 & 31.9 & 32.5 \\
LA19 cond. (strata) & 24.7 & \textbf{21.9} & 25.3 & 26.9 \\
SpoofCeleb (strata) & \textbf{23.5} & 28.0 & 25.9 & 28.3 \\
STOPA (strata) & \textbf{17.4} & 18.2 & 18.0 & 23.0 \\
Mean & 24.7 & 26.0 & 26.4 & 28.7 \\
\midrule
\multicolumn{5}{@{}l}{\textit{Detection}} \\
ASVspoof 2021 DF & 4.2 & 2.3 & \textbf{0.8} & \textbf{0.8} \\
ASVspoof 2021 LA & 8.7 & 10.2 & \textbf{4.1} & 7.0 \\
ASVspoof 5 & 10.4 & \textbf{9.6} & 11.4 & 11.8 \\
CodecFake+ CoSG & 14.9 & 12.4 & 6.8 & \textbf{6.3} \\
CodecFake+ & \textbf{1.3} & 3.2 & 3.6 & 4.5 \\
DFADD & \textbf{0.0} & 0.1 & \textbf{0.0} & 0.2 \\
DFADD aug. & \textbf{34.3} & 35.4 & 36.4 & 35.9 \\
DFADD cond. & \textbf{30.8} & 31.1 & 34.0 & 32.3 \\
In-the-Wild & 6.1 & 4.8 & \textbf{2.7} & 4.0 \\
ASVspoof 2019 LA & 0.6 & \textbf{0.5} & 0.7 & 1.2 \\
LA19 aug. & \textbf{32.4} & 36.5 & 33.3 & 33.6 \\
LA19 cond. & \textbf{27.4} & 30.4 & 29.8 & 29.4 \\
LibriSeVoc & \textbf{0.0} & 0.3 & \textbf{0.0} & \textbf{0.0} \\
LibriSeVoc aug. & 35.6 & 35.4 & 26.0 & \textbf{25.0} \\
LibriSeVoc cond. & 27.3 & 27.0 & \textbf{22.9} & 23.3 \\
SpoofCeleb & 0.3 & 1.0 & \textbf{0.1} & 0.8 \\
Mean & 14.7 & 15.0 & 13.3 & 13.5 \\
\bottomrule
\end{tabular}
\end{adjustbox}
\end{table}

%% file: tables/app_examples.tex
\begin{figure}[!htbp]
\centering\small
\begin{tcolorbox}[colback=white, colframe=muted, boxrule=0.5pt, arc=2pt, left=6pt, right=6pt, title={\small Detection
rationale, PartialEdit (one word edited, labelled span 1.54--2.05\,s); one-step answer: genuine, $P(\text{spoof})=0.09$},
coltitle=ink, colbacktitle=oursbg, fonttitle=\small]
\ttfamily\footnotesize
OBSERVATIONS: Reduced bandwidth (about 5-7 kHz); SNR about 30 dB; speech level about -30 dBFS; no clipping.\\
LOCALIZATION: Synthesized speech at 1.60-2.24 s; the rest is genuine speech. The 3.4 s recording repeats in the
window; times refer to its first occurrence.\\
SOURCE\_PROFILE: The generation method cannot be determined.\\
DECISION: The recording is a spoof: synthetic segments are embedded in genuine speech.\\[2pt]
\rmfamily\small Checks: bandwidth fails (measured full band); SNR and level hold; span IoU 0.64; decision correct.
\end{tcolorbox}
\begin{tcolorbox}[colback=white, colframe=muted, boxrule=0.5pt, arc=2pt, left=6pt, right=6pt, title={\small Verification
rationale, STOPA REVERSAL pair (different systems, every attribute class shared)}, coltitle=ink, colbacktitle=oursbg,
fonttitle=\small]
\ttfamily\footnotesize
CHANNEL\_STATUS: The first recording has no audible noise, no noticeable reverberation and Opus compression; the second
recording has no audible noise, no noticeable reverberation and no codec compression.\\
FINGERPRINT: The latent source-fingerprint cues show clear differences.\\
PROFILE\_COMPATIBILITY: Generation method is similar (both copy-synthesis of real speech). Acoustic modelling is similar
(both copy-synthesis without an acoustic model). Waveform generation differs (time-domain neural-codec decoders vs GAN
vocoders).\\
DECISION: Channel degradations are present, but they are not evidence about the generation source. Across the source
profile, the attributes are mixed. This agrees with the fingerprint cues. Taken together, the recordings come from
different generation systems.\\[2pt]
\rmfamily\small Checks: decision correct; the attribute statements fail (both systems are diffusion / flow-matching TTS
with GAN vocoders), which the check exposes while the fingerprint verdict carries the decision.
\end{tcolorbox}
\caption{Verbatim outputs of \name{}-Base with their checks.}
\label{fig:app-examples}
\end{figure}